\pdfoutput=1
\documentclass[conference]{IEEEtran}

\usepackage[available,functional,reproduced]{ndssbadges}
\usepackage{tikz}
\usetikzlibrary{positioning, arrows.meta}
\usepackage{amsmath}
\usepackage{amssymb}
\usepackage{array}
\usepackage{booktabs}
\usepackage{graphicx}
\usepackage{multirow}
\usepackage{makecell}
\usepackage{siunitx}
\usepackage{stfloats}
\usepackage{url}
\usepackage[hidelinks]{hyperref}
\usepackage{cite}


\ifCLASSOPTIONcompsoc
  \usepackage[caption=false,font=normalsize,labelfont=sf,textfont=sf]{subfig}
\else
  \usepackage[caption=false,font=footnotesize]{subfig}
\fi

\newcommand{\res}[2]{%
  #1~{\color{black!60}\footnotesize $\pm$~#2}%
}

\newcommand{\bres}[2]{%
  \textbf{#1}~{\color{black!60}\footnotesize $\pm$~#2}%
}

\usepackage[table]{xcolor}
\usepackage{pgf}

\newcommand{\Description}[1]{}

\newcommand{\maxscale}{0.8}  
\newcommand{\maxfpr}{0.01}   
\newcommand{\minauroc}{0.5}  
\newcommand{\maxauroc}{1.0}  

\newcommand{\applygradient}[1]{%
  \pgfmathsetmacro{\calculatedopacity}{max(0,min(25,#1/\maxscale*20))}%
  \edef\temp{\noexpand\cellcolor{blue!\calculatedopacity}}\temp%
}

\newcommand{\applyaurocgradient}[1]{%
  \pgfmathsetmacro{\rawratio}{%
    (#1-\minauroc)/(\maxauroc-\minauroc)%
  }%
  \pgfmathsetmacro{\scaledratio}{max(0,min(1,\rawratio))}%
  \pgfmathsetmacro{\calculatedopacity}{\scaledratio*20}%
  \edef\temp{\noexpand\cellcolor{blue!\calculatedopacity}}\temp%
}

\newcommand{\applyinversegradient}[1]{%
  \pgfmathsetmacro{\rawratio}{#1/\maxfpr}%
  \ifdim\rawratio pt>1pt
    \def\rawratio{1}%
  \fi%
  \pgfmathsetmacro{\scaledratio}{sqrt(\rawratio)}%
  \pgfmathsetmacro{\calculatedopacity}{(1-\scaledratio)*20}%
  \edef\temp{\noexpand\cellcolor{blue!\calculatedopacity}}\temp%
}

\newcommand{\gres}[2]{\applygradient{#1}\res{#1}{#2}}
\newcommand{\gbres}[2]{\applygradient{#1}\bres{#1}{#2}}

\newcommand{\gaucres}[2]{\applyaurocgradient{#1}\res{#1}{#2}}
\newcommand{\gaucbres}[2]{\applyaurocgradient{#1}\bres{#1}{#2}}

\newcommand{\gfpres}[2]{\applyinversegradient{#1}\res{#1}{#2}}
\newcommand{\gfpbres}[2]{\applyinversegradient{#1}\bres{#1}{#2}}

\newcommand{\yesmark}{%
  \cellcolor{green!12}\textcolor{green!35!black}{\textbf{Yes}}%
}
\newcommand{\nomark}{%
  \cellcolor{gray!12}\textcolor{gray!60!black}{\textbf{No}}%
}
\newcommand{\variesmark}{%
  \cellcolor{orange!15}\textcolor{orange!40!black}{\textbf{Varies}}%
}

\hypersetup{
  pdftitle={How Benchmarks and Evaluation Protocols Shape Conclusions in Provenance-Based Intrusion Detection},
  pdfauthor={Lorenzo Guerra, Thomas Chapuis, Guillaume Duc, Pavlo Mozharovskyi, and Van-Tam Nguyen}
}

\title{How Benchmarks and Evaluation Protocols\\
Shape Conclusions in Provenance-Based\\
Intrusion Detection}

\newcommand{\affmark}[1]{\textsuperscript{#1}}

\author{
\IEEEauthorblockN{
Lorenzo Guerra\affmark{1,2},
Thomas Chapuis\affmark{2},
Guillaume Duc\affmark{1},
Pavlo Mozharovskyi\affmark{1},
Van-Tam Nguyen\affmark{1}
}

\IEEEauthorblockA{
\affmark{1}LTCI, T\'el\'ecom Paris, Institut Polytechnique de Paris\\
\{lorenzo.guerra, guillaume.duc, pavlo.mozharovskyi, van-tam.nguyen\}@telecom-paris.fr
}

\IEEEauthorblockA{
\affmark{2}Ampere Software Technology\\
\{lorenzo.guerra, thomas.chapuis\}@ampere.cars
}
}

\begin{document}

\IEEEoverridecommandlockouts
\makeatletter\def\@IEEEpubidpullup{6.5\baselineskip}\makeatother
\IEEEpubid{\parbox{\columnwidth}{
    Network and Distributed System Security (NDSS) Symposium 2027\\
    22--26 March 2027, Seoul, Republic of Korea\\
    ISBN 978-1-970672-09-1\\
    https://dx.doi.org/10.14722/ndss.2027.230448\\
    www.ndss-symposium.org
}
\hspace{\columnsep}\makebox[\columnwidth]{}}

\maketitle

\begin{abstract}
Provenance-based intrusion detection systems (PIDS) frequently report strong performance, but the conclusions drawn from these results can be highly sensitive to benchmarking choices and evaluation protocols. We investigate this dependency by re-evaluating representative PIDS on public datasets that meet our audit, labeling, and calibration requirements. Focusing primarily on the audited DARPA TC E3 datasets, we apply a unified protocol with temporally separated test periods and validation-only checkpoint selection and threshold calibration, and ask which architectural claims are empirically supported. We find that alerting success and investigation utility can diverge sharply, as several systems surface attacks without providing enough process-level context to support forensic investigation. Across the four primary datasets, a simple allowlist built from executable names and paths observed during training matches or exceeds the selected learned baselines on key operating-point metrics, showing that comparable performance on these metrics is achievable using lexical novelty alone. Quantifying semantic signal quality through feature completeness and field entropy helps explain why several audited E3 datasets support alerting performance without reliably separating model architectures. Theia, however, provides the richest semantic signal and is the only primary dataset on which our reference model, Theseus, clearly improves both ranking and node-level recovery over the other learned systems. Overall, these findings reinforce the importance of interpreting architectural claims in PIDS together with the benchmark properties and evaluation protocol that produced them.
\end{abstract}

\IEEEpeerreviewmaketitle

\section{Introduction}\label{sec:intro}

Provenance-based intrusion detection systems (PIDS) are widely studied for host-based security~\cite{zipperle2022provenance,dong2023we}. They represent system execution as a directed graph of causal relations between entities such as processes, files, and network connections, allowing multi-step attacks to be analyzed beyond individual log events~\cite{zipperle2022provenance,sleuth,holmes}. Graph neural networks (GNNs) have become a common modeling choice for this setting because they can encode the structural relationships present in provenance graphs~\cite{bilot2023gnn,kairos,threatrace}.

Reported detection results, however, are difficult to compare across papers. Prior studies use different benchmarks, labeling strategies, data splits, calibration rules, and reported metrics~\cite{abrar2025reproducibility,bilot2025simpler,liu2025what}. Bilot et al.~\cite{bilot2025simpler} improve comparability by evaluating multiple PIDS baselines in a shared framework and exposing several evaluation distortions. Such breadth is valuable, but architectural conclusions also depend on whether the benchmark preserves enough usable signal to distinguish richer models from simpler alternatives. In provenance benchmarks, graph construction and collection artifacts can change the available signal, so measured performance may reflect shortcut cues or dataset artifacts rather than architectural differences~\cite{michael2020forensic,geirhos2020shortcut}.

Evaluation protocols introduce another source of ambiguity. Alerting metrics such as Attack Detection Precision (ADP)~\cite{bilot2025simpler} evaluate whether a system can surface at least one useful signal from an attack, which is important for triage. However, this does not address the forensic task of ranking and isolating attack-relevant entities from large volumes of benign activity~\cite{jiang2025orthrus,hassan2019nodoze,fang2022back,li2024nodlink}. We report both alerting and investigation metrics, covering ranking quality and node-level recovery. We also enforce validation-only calibration, with both checkpoints and thresholds selected from validation data, since selecting either on the test period gives the detector information that would not be available before deployment~\cite{arp2022and,bilot2025simpler,abrar2025reproducibility}. The resulting protocol keeps labels, artifact handling, and calibration fixed across systems, with performance reported over multiple random seeds.

We apply this methodology to public provenance datasets and use the DARPA TC E3 family~\cite{darpa-tc-e3} as the primary setting. E3 is the most suitable public setting we found for this controlled comparison because it provides reproducible process-level labels, publicly documented artifact analysis, and temporal structure compatible with validation-only calibration. We refer to the E3 subset evaluated using the REAPr labels~\cite{reaprgroundtruth} and Liu et al.'s artifact analysis and scoring exclusions~\cite{liu2025what} as \emph{audited E3}. We report ATLASv2~\cite{riddle2023atlasv2} separately because its provenance graph, reconstructed from EDR telemetry, conservatively overapproximates possible dependencies~\cite{liu2026windows}. Its attacks are concentrated in a single period and share substantial payload structure, leaving no independent attack data for validation.

Within this scope, we examine which architectural conclusions survive a common protocol for calibration and scoring. The audited E3 datasets retain detectable attack signal, but most provide limited evidence for comparing learned architectures or evaluating process-level investigation. On Cadets, FiveDirections, and Trace, an allowlist over executable names and paths observed during training achieves competitive fixed-threshold performance, showing that these results alone do not establish the benefit of learned provenance representations. To evaluate whether greater model capacity helps when richer signal is available, we introduce Theseus, a reference model that combines E-GraphSAGE with a Transformer autoencoder over windowed provenance graphs to produce process-level anomaly scores. Theseus shows no clear advantage over simpler baselines on Cadets, FiveDirections, or Trace, consistent with the allowlist diagnostic. Theia is the exception, with more complete and varied semantic fields, and there Theseus improves both anomaly ranking and node-level recovery over the learned baselines. These results suggest that the benefits of richer modeling depend on the semantic information available in the evaluated dataset.

The paper makes the following contributions:
\begin{itemize}
\setlength\itemsep{0em}
\item We present an auditable PIDS evaluation protocol that combines shared label scope, artifact handling, temporal splitting, validation-only checkpoint selection and threshold calibration, and multi-seed reporting with metrics for both alerting and investigation utility.
\item We re-evaluate representative baselines under this shared methodology and show that high precision and ADP do not necessarily imply strong node-level recovery.
\item We analyze whether provenance datasets support architectural comparison by measuring semantic signal quality through feature completeness and field entropy. This analysis helps explain why some datasets reward lexical novelty or structural heuristics more than richer provenance modeling.
\item We introduce Theseus as a reference model for testing when additional model capacity improves provenance-based detection. In our audited E3 evaluation, its clearest gains occur on Theia, which has the highest semantic signal quality.
\end{itemize}

\section{Related Work}

\subsection{Provenance Collection and Reduction}
PIDS represent system activity as causal graphs. Provenance frameworks such as SPADE~\cite{spade}, CamFlow~\cite{camflow}, and Linux Provenance Modules~\cite{bates2015whole-system} showed how system-level provenance records can be represented as directed graphs of entities, including processes, files, and network connections, together with their causal interactions. Later systems used these graphs directly for security analysis. SLEUTH~\cite{sleuth} reconstructs attack scenarios through tag propagation, while Holmes~\cite{holmes} correlates suspicious information flows to detect advanced persistent threats. A recurring challenge is dependence explosion, where long-running processes accumulate dense and noisy causal histories. This limits the scalability of raw provenance graphs for detection and investigation, motivating provenance-reduction techniques~\cite{lee2013loggc,beep,xu2016reduction} and priority-driven causality analysis~\cite{liu2018towards}.

\subsection{Learning on Provenance Graphs}
Recent work on intrusion detection has used learned, semantic, and graph-derived representations to reduce manual feature engineering over large audit streams~\cite{zeng2021watson,bilot2023gnn}. Sequential log models such as DeepLog~\cite{deeplog} and LogBERT~\cite{guo2021logbert} apply neural anomaly detection to system logs without explicitly representing causal dependencies among system entities, while provenance-based detectors use these dependencies to provide context for detection. Unicorn~\cite{unicorn} summarizes long-running provenance graphs with streaming sketches, while ProvDetector~\cite{wang2020provdetector} selects suspicious provenance paths and embeds them for outlier detection. More recent systems learn representations directly on graphs. Kairos~\cite{kairos} uses a GNN encoder--decoder to model temporal changes in provenance graphs and assign anomaly scores to events. Magic~\cite{jia2024magic} uses masked graph representation learning, while ThreaTrace~\cite{threatrace} and Flash~\cite{flash} compute node or neighborhood representations to identify suspicious graph elements. R-CAID~\cite{rcaid} takes a different direction by embedding root-cause information into learned provenance-based detection.

These systems often report strong detection results, but their empirical claims are difficult to compare directly because they use different label scopes, temporal splits, thresholding policies, and reporting conventions~\cite{abrar2025reproducibility,bilot2025simpler,liu2025what}.

\subsection{PIDS Evaluation and Benchmark Validity}

Recent work has examined PIDS evaluation more directly. Abrar et al.~\cite{abrar2025reproducibility} audited the reproducibility of deep-learning-based PIDS and found that incomplete code, missing documentation, unavailable data, missing preprocessing steps, and unclear experimental procedures limit independent verification. Bilot et al.~\cite{bilot2025simpler} improved comparability by implementing multiple PIDS baselines in a unified framework, introducing ADP as an alerting metric, and showing that a lightweight detector, Velox, can match or exceed more complex systems across many DARPA datasets. Liu et al.~\cite{liu2025what} shifted attention to dataset quality, showing that benchmark contents, background activity, collection artifacts, and attack conspicuousness can substantially affect endpoint threat-detection results.

Our work connects these views. Broad baseline comparisons show that simple PIDS models can match or exceed richer architectures under many published benchmark settings, raising the question of whether those results reflect model design or the benchmark and protocol used to measure it. Dataset-quality studies show that artifacts, background structure, and label ambiguity can shape measured performance, but they do not by themselves explain when a benchmark can separate learned architectures under a shared protocol. We combine these perspectives in a controlled evaluation that asks which claims about model design remain visible once labels, artifact handling, calibration, and scoring are fixed.

Compared with prior broad evaluations, our protocol builds on Liu et al.'s artifact audit~\cite{liu2025what}, extends validation-based selection to checkpoints as well as thresholds, and reports both alerting behavior and investigation utility.

This setup lets us ask whether the benchmark exposes enough usable signal for richer modeling. Dataset content dominated by sparsity, repetition, or artifacts may leave too little signal to distinguish richer architectures from simpler alternatives, while more complete and diverse semantic attributes can make those differences observable under the same protocol. We use semantic signal quality to characterize the available semantic information and help explain why architectural differences are more apparent on some datasets than others.

\section{Evaluation Protocol Distortions}

While prior PIDS evaluations vary widely, several protocol choices can inflate reported performance or mask critical failure modes~\cite{arp2022and,bilot2025simpler}. We focus on three controls needed for architectural comparison in our study: separating alerting from investigation utility, preventing test-period information from influencing calibration, and accounting for initialization variance. The first two affect what conclusions a benchmark can support, while the third prevents stochastic training effects from being mistaken for architectural differences.

\subsection{Alerting and Investigation Metrics}
No single metric captures both the alerting and forensic roles of a provenance-based detector. Bilot et al.~\cite{bilot2025simpler} introduced Attack Detection Precision (ADP), which credits a system for surfacing at least one true positive from each distinct attack scenario without requiring full recovery of the attack footprint. This matches an important SOC use case, as one reliable alert can be enough to start triage.

However, a single alert provides limited evidence of the detector’s usefulness for investigation. Analysts must reconstruct the malicious footprint and distinguish it from large volumes of benign activity~\cite{fang2022back,hassan2019nodoze,liu2018towards,li2024nodlink}. As Jiang et al.~\cite{jiang2025orthrus} argue through the notion of ``Quality of Attribution'' (QoA), an alert is less useful if it leaves analysts to manually recover the relevant nodes from a massive graph. ADP captures whether a system surfaces at least one signal from an attack, but not how well it ranks or recovers the remaining attack-relevant entities. We pair ADP with average precision (AP), area under the receiver operating characteristic curve (AUROC), and MCC. We use standard average precision, computed as $\sum_n (R_n-R_{n-1})P_n$, where $P_n$ and $R_n$ denote precision and recall at successive score thresholds. Because labeled attack processes are rare, AP remains our primary ranking metric, as it reflects precision among retrieved processes, whereas AUROC can remain high under severe imbalance even when precision at useful operating points is poor~\cite{davis2006relationship,saito2015precision,arp2022and}. MCC complements both ranking metrics by evaluating the selected operating point and penalizing missed attacks and false positives~\cite{chicco2020advantages}.

\subsection{Temporal Leakage and Calibration}
A second distortion comes from weak separation between model development and final testing~\cite{arp2022and,bilot2025simpler}. Using final evaluation data for model selection, threshold calibration, or post-processing gives the detector information that would not be available before deployment, which is a form of data snooping~\cite{arp2022and}. Bilot et al.~\cite{bilot2025simpler} identify the same issue in PIDS evaluations, including thresholding and clustering techniques that depend on the full score distribution or assume that malicious outliers are present.

Random splits create a related problem by mixing future behavior into training or validation data. Similar concerns about temporal bias have been observed in evaluation for intrusion detection and malware detection, where unrealistic train/test splits can hide the distribution shifts faced after deployment~\cite{kan2025tesseract,pragmatic-assessment}. Selecting checkpoints on the test period tunes the model to that period, while random splitting masks the temporal structure that a deployed detector must handle.

Our evaluation keeps checkpoint selection and threshold calibration independent of the held-out period, preventing these choices from adapting to the evaluation data.

\subsection{Initialization Variance}
PIDS models can vary substantially across random seeds~\cite{bilot2025simpler}. Similar variance has been observed across machine-learning benchmarks more broadly, where stochastic training choices can make single-run comparisons unreliable~\cite{reimers2017reporting,bouthillier2021accounting}. For this reason, we treat multi-seed reporting as part of the evaluation protocol and report mean performance together with standard deviation, so that a favorable initialization is not mistaken for an architectural effect.

\section{Methodology}\label{sec:methodology}

\subsection{Benchmark Selection and Preparation}

\begin{table}[t]
    \centering
    \small
    \setlength{\tabcolsep}{3pt}
    \caption{Core viability criteria for the primary regime.}
    \label{tab:benchmark_viability_main}
    \begin{tabular}{lcccl}
        \toprule
        \makecell[l]{\textbf{Benchmark}} &
        \makecell[c]{\textbf{Compatible}\\\textbf{labels}} &
        \makecell[c]{\textbf{Dataset audit}\\\textbf{support}} &
        \makecell[c]{\textbf{Attack}\\\textbf{validation}} &
        \makecell[l]{\textbf{Role}} \\
        \midrule
        Audited E3 & \yesmark & \yesmark & \yesmark & Primary \\
        ATLASv2 & \yesmark & \yesmark & \nomark & Secondary \\
        E5 & \nomark & \nomark & \variesmark & Excluded \\
        OpTC & \nomark & \yesmark & \nomark & Excluded \\
        \bottomrule
    \end{tabular}
\end{table}

\subsubsection{Primary E3 Regime}
We evaluate models on an audited subset of the DARPA TC E3 provenance datasets~\cite{darpa-tc-e3}: Cadets, FiveDirections, Theia, and Trace. These datasets were collected during the same DARPA engagement and share the Common Data Model (CDM) representation, but they span different hosts, attack strategies, and operating systems. This gives us a controlled setting in which the main sources of evaluation variation can be fixed across datasets.

Our primary regime requires three conditions. The dataset must provide reproducible process-level ground truth under a shared labeling procedure, public audit support characterizing collection artifacts and ambiguous nodes, and enough temporal structure to isolate the test set and support validation-only calibration. Audited E3 satisfies these requirements through REAPr labels~\cite{reaprgroundtruth} and Liu et al.'s artifact analysis~\cite{liu2025what}, which together define the scoring target and identify contaminated nodes to exclude. These constraints reduce cases where label construction, artifact handling, or threshold selection explain a performance difference. Table~\ref{tab:benchmark_viability_main} summarizes the resulting benchmark roles, and Appendix~\ref{app:benchmark_viability} reports the selection details for each dataset.

\subsubsection{Secondary ATLASv2 Evaluation}
\label{sec:atlas_methodology}
\begin{table*}[!t]
    \centering
    \small
    \sisetup{group-separator={,}, group-minimum-digits=4}
    \caption{ATLASv2 graph adaptation and audit of UUID-based process labels. Event counts are reported separately for training and testing. Attack prevalence is the percentage of scored test processes labeled as attack. Match rate is the percentage of REAPr label rows mapped uniquely to reconstructed process nodes through Carbon Black process UUIDs.}
    \label{tab:atlas_audit}
    \begin{tabular}{l
                    S[table-format=3.0]
                    S[table-format=3.0]
                    S[table-format=4.0, group-separator={,}]
                    S[table-format=2.0]
                    S[table-format=3.0]
                    S[table-format=1.2]
                    S[table-format=3.1]}
    \toprule
    \textbf{Dataset} &
    {\makecell[c]{\textbf{Train}\\\textbf{Events (k)}}} &
    {\makecell[c]{\textbf{Test}\\\textbf{Events (k)}}} &
    {\makecell[c]{\textbf{Test}\\\textbf{Processes}}} &
    {\makecell[c]{\textbf{Attack}\\\textbf{Scenarios}}} &
    {\makecell[c]{\textbf{Attack}\\\textbf{Processes}}} &
    {\makecell[c]{\textbf{Attack}\\\textbf{Prevalence (\%)}}} &
    {\makecell[c]{\textbf{Match}\\\textbf{Rate (\%)}}} \\
    \midrule
    ATLASv2-h1 & 945 & 326 & 7712 & 10 & 161 & 2.09 & 100.0 \\
    ATLASv2-h2 & 492 & 186 & 7413 & 6  & 30  & 0.40 & 100.0 \\
    \bottomrule
    \end{tabular}
\end{table*}

ATLASv2~\cite{atlas,riddle2023atlasv2} provides a secondary evaluation setting with public Windows endpoint telemetry, but it does not match the audited E3 setting directly. We reconstruct a process-centric provenance graph from the Carbon Black EDR telemetry, which provides stable process GUIDs together with broad coverage of process, file, and network activity. Unlike the CDM audit streams used in E3, however, the EDR telemetry records the start of a data access rather than individual read and write operations, and it lacks stop events for data flows. The reconstructed graph therefore conservatively overapproximates possible dependencies~\cite{liu2026windows}.

We use the UUID-based REAPr ATLASv2 process labels, which expose the Carbon Black process UUID for every labeled attack process and define the attack scope using Windows EDR tracing heuristics~\cite{reaprgroundtruth,liu2026windows}. All attack processes labeled in h1 and h2 map uniquely to reconstructed process nodes. ATLASv2 nevertheless lacks an independent validation split containing attack data, as each host contains only one engagement window with attack activity and the scenarios share substantial payload structure~\cite{riddle2023atlasv2}. Given this limitation, we calibrate thresholds from benign training scores and report ATLASv2 separately as corroborative evidence. Table~\ref{tab:atlas_audit} summarizes the resulting adaptation.

\subsubsection{Excluded Benchmarks}
E5 and OpTC fall outside the primary protocol because neither satisfies all three viability criteria. E5 has been used in prior broad evaluations, but we are not aware of public audit support that documents artifact handling, ambiguous node treatment, and scoring exclusions for the label target used in those evaluations~\cite{jiang2025orthrus,bilot2025simpler}. OpTC has stronger public audit support after the correction and labeling work of Majorczyk et al.~\cite{majorczyk2025newhope}, but its labels reconstruct malicious host events and network flows from OpTC-specific evidence rather than the REAPr process-level target used here~\cite{reaprgroundtruth,liu2025what}. It also lacks an independent validation split containing attack data. We keep the main comparison on audited E3 and leave corrected OpTC to a separate evaluation with its own graph construction and label reconciliation. Appendix~\ref{app:benchmark_viability} provides the dataset-specific details.

\subsubsection{E3 Artifact Handling}
Within E3, we follow the artifact analysis protocol of Liu et al.~\cite{liu2025what}. Confirmed collection errors are removed from scoring and calibration, while ordinary system noise is retained. For Cadets, we keep the \path{wwtawwtal_bad_neighborhood} artifact in the processed graphs and model inference, but exclude its labeled nodes from threshold calibration and all reported metric computation. For Theia, we use only the final stable time slice, T3, to avoid artifacts caused by changes to the workload generator’s configuration~\cite{liu2025what}. For Trace, we use a contiguous five-day window around the attack, which keeps computation tractable while preserving benign diversity. Table~\ref{tab:dataset_stats} reports aggregate statistics for the primary E3 temporal graph snapshots produced by our pipeline.

\begin{table*}[t]
    \centering
    \small
    \caption{E3 graph statistics across Theseus training, validation, and test snapshots. Node and edge counts include repeated instances across snapshots, while unique entities are counted once. P/F/N denotes percentages of process, file, and network flow node instances.}
    \label{tab:dataset_stats}
    \begin{tabular}{l
                    S[table-format=4.0, group-separator={,}]
                    S[table-format=3.0, group-separator={,}]
                    S[table-format=3.0, group-separator={,}]
                    S[table-format=3.0, group-separator={,}]
                    S[table-format=7.0, group-separator={,}]
                    S[table-format=7.0, group-separator={,}]
                    S[table-format=8.0, group-separator={,}]
                    S[table-format=4.0, group-separator={,}]
                    S[table-format=5.0, group-separator={,}]
                    c}
        \toprule
        \textbf{Dataset} &
        {\makecell[c]{\textbf{Total}\\\textbf{Snapshots}}} &
        {\makecell[c]{\textbf{Train}\\\textbf{Snap.}}} &
        {\makecell[c]{\textbf{Val.}\\\textbf{Snap.}}} &
        {\makecell[c]{\textbf{Test}\\\textbf{Snap.}}} &
        {\makecell[c]{\textbf{Unique}\\\textbf{Entities}}} &
        {\makecell[c]{\textbf{Node}\\\textbf{Instances}}} &
        {\makecell[c]{\textbf{Edge}\\\textbf{Instances}}} &
        {\makecell[c]{\textbf{Median}\\\textbf{Nodes}}} &
        {\makecell[c]{\textbf{Median}\\\textbf{Edges}}} &
        \makecell[c]{\textbf{Distribution}\\\textbf{(\textit{P/F/N})}} \\
        \midrule
        Cadets & 1100 & 634 & 196 & 270 & 658429 & 1319690 & 9656378 & 524 & 7518 & 18\% / 75\% / 7\% \\
        FiveDir. & 1168 & 445 & 253 & 470 & 868430 & 5257253 & 15316414 & 5441 & 10972 & 6\% / 91\% / 3\% \\
        Trace & 635 & 314 & 114 & 207 & 2635286 & 3494952 & 8434106 & 4926 & 7456 & 15\% / 34\% / 51\% \\
        Theia (T3) & 390 & 117 & 55 & 218 & 892664 & 1737117 & 11052366 & 4367 & 14300 & 10\% / 67\% / 23\% \\
        \bottomrule
    \end{tabular}
\end{table*}

\subsection{Temporal Splits and Calibration}
\label{sec:temporal_calibration}

The held-out test period is never used for training, checkpoint selection, or threshold calibration, preventing these choices from using future evaluation data~\cite{abrar2025reproducibility,arp2022and}.

Within the development period, a strictly monotonic \textit{train $\rightarrow$ validate $\rightarrow$ test} schedule is not always practical after artifact treatment, because it would either leave too little attack data for reliable checkpoint selection or discard useful benign history. We draw both training and validation data from the pre-test pool while reserving the final period only for evaluation. Validation attacks are used only to select checkpoints by validation AP, since validation loss can continue decreasing even after detection performance has peaked. This mismatch arises because the self-supervised training objective is not perfectly aligned with the downstream anomaly detection task. Threshold calibration uses only benign validation scores, with the deployment threshold set to the maximum benign validation anomaly score.

We retain entities observed during the training period even if they are labeled as malicious later. E3 labels can identify a process involved in an attack without specifying when its malicious behavior began, so the same process may also appear in earlier benign activity. Excluding it from training based on its later attack label would introduce knowledge from the test period and remove activity the detector would otherwise have observed. This could make the process appear artificially novel during testing and inflate its anomaly score.

\subsection{Scoring Scope and Ground Truth}
Our main evaluation uses the process-level labels released in REAPr~\cite{reaprgroundtruth}, together with the artifact analysis and dataset audit of Liu et al.~\cite{liu2025what}. These labels define the scoring scope for all main metrics. Processes in the scored attack chain are identified by intersecting a forward trace from the attack's root causes with a backward trace from its terminal impacts. Processes reached by the forward trace but not the backward trace are labeled as ``contaminated'' and excluded from scoring. These processes may exhibit anomalous behavior caused by the attacker, but DARPA's attack engagement reports describe the intrusions only at a coarse procedural level~\cite{darpa-tc-e3}. Excluding them avoids rewarding or penalizing detections whose connection to the documented attack chain cannot be established confidently.

Metrics are computed only over process nodes. File and network flow nodes remain available during training and message passing but are excluded from scoring, since treating these unlabeled node types as benign would assign labels unsupported by the ground truth. Section~\ref{sec:label_granularity} discusses the limits of process-level scoring and how finer annotations can help identify specific malicious actions.

\subsection{Semantic Signal Quality Diagnostic}
\label{sec:semantic_signal}

We characterize the semantic information available to richer models using a diagnostic based on feature completeness and field entropy in the processed node tables. Completeness measures how often predefined semantic attributes are populated, while entropy measures token-level diversity among populated values. The diagnostic helps interpret performance differences across datasets and is not used for training, model selection, or threshold calibration.

Let $\mathcal{F}$ denote the set of applicable semantic fields in the processed node tables, including process path, process command line, file path, and network flow description. The network flow field is treated as populated when its source address, source port, destination address, and destination port are all present. Feature completeness is
\begin{equation}
    C = \frac{\sum_{f \in \mathcal{F}} \text{filled}_f}
             {\sum_{f \in \mathcal{F}} \text{total}_f},
\end{equation}
where $\text{filled}_f$ is the number of populated records for field $f$ under the criteria above and $\text{total}_f$ is the number of records to which that field applies.

We compute token frequencies using the same field-specific tokenization rules as Theseus, without applying embedding weights. For each field $f$, let $h_f$ be the Shannon entropy~\cite{shannon1948mathematical} of the distribution of tokens across all populated values in the corpus, measured in bits. We aggregate field-level entropy as
\begin{equation}
    H = \frac{\sum_{f \in \mathcal{F}} h_f \cdot \text{filled}_f}
             {\sum_{f \in \mathcal{F}} \text{filled}_f},
\end{equation}
which averages entropy values in proportion to the number of populated records. Missing values are excluded from this denominator because sparsity is captured by $C$.

Semantic signal quality combines these quantities as
\begin{equation}
    Q = C \cdot H.
\end{equation}
The product is lower when semantic fields are sparsely populated or their token distributions are concentrated on a small vocabulary. We interpret $Q$ as a descriptive diagnostic within the audited benchmarks, alongside the broader audit in Section~\ref{sec:benchmark_guidance}, which examines entity linkage, raw feature coverage, temporal continuity, and label granularity. Section~\ref{sec:ssq_analysis} reports the values for each dataset.

\subsection{Baselines and Diagnostic Allowlist}

We select representative baselines with strong reported performance and reproducible implementations that can be adapted to our shared protocol without reimplementing their core model logic. This selection is informed by recent work on the reproducibility of PIDS based on deep learning~\cite{abrar2025reproducibility}.

The selected systems are Velox~\cite{bilot2025simpler}, a lightweight MLP-based anomaly detector; Orthrus~\cite{jiang2025orthrus}, which combines attention-based temporal graph learning with self-supervised edge type prediction; and Magic~\cite{jia2024magic}, which uses self-supervised masked graph representation learning. We adapt each baseline to our shared calibration and label definitions while keeping their officially released model logic intact.

We also include a deterministic diagnostic allowlist that flags any process whose executable name or path was unseen during training. It tests how much performance at the selected operating point can be achieved using executable and path novelty alone. It is not intended as a standalone defense, since attacks that abuse known binaries, commonly called ``Living off the Land,'' can evade this novelty rule~\cite{barr2021survivalism}.

\subsection{Theseus: A Reference Model}\label{sec:theseus}
\begin{figure}[t]
  \centering
  \includegraphics[width=\columnwidth]{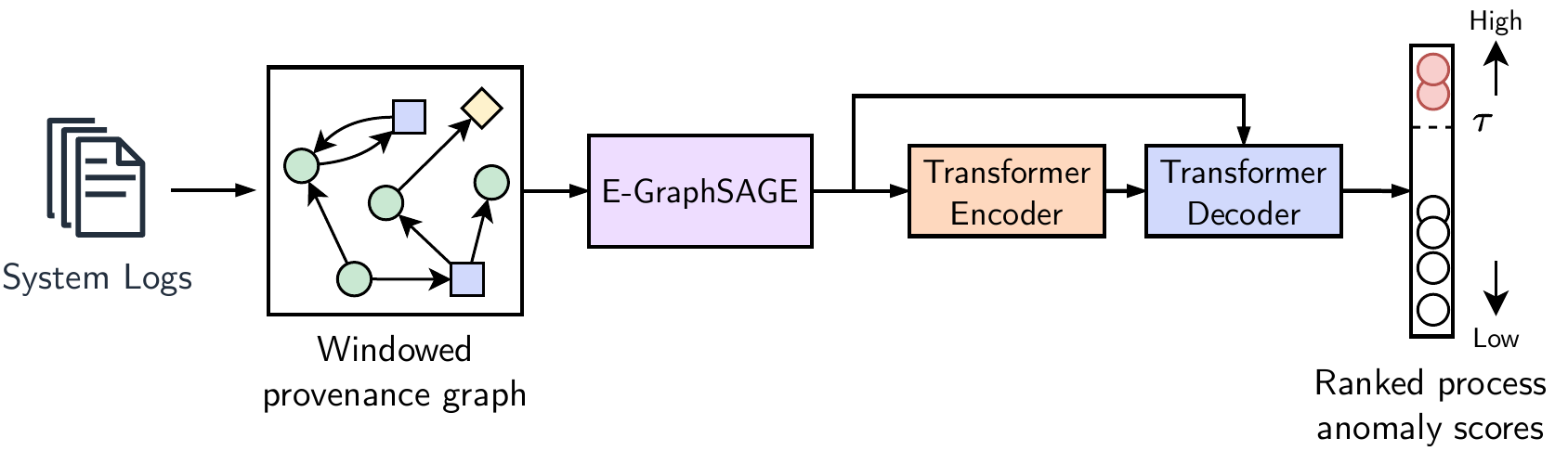}
  \caption{Theseus pipeline. System logs are converted into windowed provenance graphs, encoded with E-GraphSAGE, and scored by a Transformer autoencoder.}\label{fig:pipeline}
\end{figure}

We introduce Theseus as a reference model to examine whether combining semantic attributes with local graph context and interactions within temporal windows improves detection across the audited datasets under the shared protocol. The architecture extends established graph-based detection frameworks~\cite{guerra2025selfsupervised} by operating on windowed heterogeneous provenance graphs and jointly encoding node and edge semantics. Theseus produces continuous node-level anomaly scores, allowing ranking quality and process recovery to be evaluated against the shared ground truth.

Theseus processes 15-minute provenance graph snapshots in two stages. First, a single-hop E-GraphSAGE encoder~\cite{lo2022egraphsage} aggregates local causal context while retaining edge attributes. Then, a Transformer autoencoder~\cite{vaswani2017attention} models co-occurrence structure within each snapshot and reconstructs the resulting node embeddings. The E-GraphSAGE update is:\begin{equation}\label{eq:gnn_aggregation}
\begin{aligned}
    \mathbf{h}_{\mathcal{N}(v)} &= \text{AGG}\left( \{ \mathbf{W}_{\text{msg}} \cdot [\mathbf{x}_u || \mathbf{r}_{uv}] \mid u \in \mathcal{N}(v) \} \right), \\
    \mathbf{z}_v &= \mathbf{W}_{\text{comb}} \cdot [\mathbf{x}_v || \mathbf{h}_{\mathcal{N}(v)}].
\end{aligned}
\end{equation}
Here, $v$ is the target node, $\mathcal{N}(v)$ is its one-hop neighborhood, $\mathbf{x}_v$ and $\mathbf{x}_u$ are the target and neighbor feature vectors, and $\mathbf{r}_{uv}$ is the attribute vector for edge $(u,v)$. The mean-pooling aggregation operator $\text{AGG}(\cdot)$ is permutation-invariant, while $\mathbf{W}_{\text{msg}}$ and $\mathbf{W}_{\text{comb}}$ are learned linear projections. The resulting embedding $\mathbf{z}_v$ is passed to the Transformer autoencoder.

We represent each snapshot as a directed heterogeneous graph $G = (V, E)$ whose nodes represent processes, files, and network flows, following prior provenance-based IDS evaluations~\cite{bilot2025simpler,jiang2025orthrus}. During graph construction, we map \texttt{EVENT\_CLONE} and \texttt{EVENT\_FORK} to a common encoding to preserve process creation edges recorded under either name. Attributes are tokenized by field type and embedded with field-specific Word2Vec models~\cite{mikolov2013efficient} trained only on the training split. Node features are formed as weighted means of these token embeddings. The token weights emphasize executable names and path components for process and file nodes, and destination service information for network flow nodes. Section~\ref{sec:ablations} examines this choice.

Because our ground truth is process-level, Theseus scores only process nodes. File and network flow nodes remain in the graph and participate in message passing, but they do not contribute directly to the reconstruction loss. For any snapshot with at least one process node, the loss is computed over $V_{\text{proc}} \subset V$:
\begin{equation}
\mathcal{L} = \frac{1}{|V_{\text{proc}}|} \sum_{v \in V_{\text{proc}}} \| \text{sg}(\mathbf{z}_v) - \hat{\mathbf{z}}_v \|^2,
\end{equation}
where $\text{sg}(\cdot)$ is the stop-gradient operator and $\hat{\mathbf{z}}_v$ is the reconstructed embedding. At evaluation time, the anomaly score for process node $v$ is the corresponding reconstruction error,
$s_v=\|\mathbf{z}_v-\hat{\mathbf{z}}_v\|^2$.

We apply a symmetric binary attention mask inside the Transformer, intended to limit memorization of exact within-window co-occurrences. Appendix~\ref{sec:mask_ratio} examines sensitivity to the mask ratio. The deployment threshold $\tau$ follows the shared policy and is set to the maximum score observed on benign validation nodes. Fig.~\ref{fig:pipeline} summarizes the pipeline.

\subsection{Implementation and Infrastructure}

Experiments ran on a compute cluster with NVIDIA L40S and A100 GPUs. Theseus can run with 80\,GB of system RAM, while baseline allocations reached 128\,GB for memory-intensive models such as Orthrus. Velox and Orthrus were run through PIDSMaker~\cite{pidsmaker-repo}, the framework released by their authors, while Magic ran on CPU using its official implementation~\cite{magic-repo}. Each baseline used an isolated software environment.

We keep the published model logic intact and adapt each codebase only as needed to run under the shared protocol. These adaptations align labels, scoring, and calibration across systems, while graph construction remains system specific.

For Orthrus, we disable clustering at evaluation time and apply the same validation-based threshold policy used for the other learned models. This follows Bilot et al.'s concern about data snooping~\cite{bilot2025simpler}, since fitting clusters to test scores uses the evaluation score distribution to determine which observations are anomalous. In deployment, attacks may be absent, and the operating point must be established without access to future test scores.

When released dataset configurations are available for a baseline, we use them as the primary settings. Otherwise, we start from the official implementation and select feasible settings within the shared resource budget. To guard against results driven by a configuration mismatch under our protocol, we also ran bounded per-dataset sensitivity sweeps around both the released and adapted baseline settings. These sweeps varied the main training, regularization, and model capacity parameters while keeping model logic, scoring target, and calibration rule fixed, and none of them changed the qualitative conclusions.

\begin{table*}[!t]
\centering
\small
\setlength{\tabcolsep}{3.5pt}
\caption{Comparative performance on the primary E3 datasets. Learned models set the threshold to the maximum benign validation anomaly score. The allowlist is deterministic. \(Q\) denotes semantic signal quality. Bold values mark the best mean.}
\label{tab:results_table}
\begin{tabular}{cl ccccccc}
\toprule
\textbf{Dataset ($Q$)}
& \textbf{System}
& \textbf{AP $\uparrow$}
& \textbf{AUROC $\uparrow$}
& \textbf{Precision $\uparrow$}
& \textbf{F1 $\uparrow$}
& \textbf{MCC $\uparrow$}
& \textbf{ADP $\uparrow$}
& \textbf{FPR $\downarrow$} \\
\midrule

\multirow{5}{*}{\shortstack[c]{Cadets\\(1.74)}}
& Allowlist
& ---
& ---
& \gres{0.429}{0.000}
& \gbres{0.021}{0.000}
& \gbres{0.058}{0.000}
& ---
& \gfpres{0.0012}{0.0000} \\

& Orthrus
& \gres{0.254}{0.009}
& \gaucbres{0.842}{0.006}
& \gres{0.600}{0.548}
& \gres{0.005}{0.004}
& \gres{0.036}{0.033}
& \gres{0.801}{0.102}
& \gfpbres{0.0000}{0.0000} \\

& Velox
& \gbres{0.266}{0.070}
& \gaucres{0.840}{0.050}
& \gbres{0.883}{0.162}
& \gres{0.007}{0.006}
& \gres{0.048}{0.022}
& \gbres{0.870}{0.031}
& \gfpres{0.0001}{0.0001} \\

& Magic
& \gres{0.159}{0.006}
& \gaucres{0.802}{0.009}
& \gres{0.000}{0.000}
& \gres{0.000}{0.000}
& \gres{0.000}{0.000}
& \gres{0.414}{0.162}
& \gfpbres{0.0000}{0.0000} \\

& \textbf{Theseus}
& \gres{0.215}{0.093}
& \gaucres{0.773}{0.142}
& \gres{0.000}{0.000}
& \gres{0.000}{0.000}
& \gres{-0.001}{0.002}
& \gres{0.400}{0.058}
& \gfpres{0.0001}{0.0001} \\

\midrule

\multirow{5}{*}{\shortstack[c]{FiveDir.\\(3.97)}}
& Allowlist
& ---
& ---
& \gbres{0.003}{0.000}
& \gbres{0.006}{0.000}
& \gbres{0.011}{0.000}
& ---
& \gfpres{0.0014}{0.0000} \\

& Orthrus
& \gbres{0.007}{0.001}
& \gaucres{0.842}{0.012}
& \gres{0.000}{0.000}
& \gres{0.000}{0.000}
& \gres{0.000}{0.000}
& \gbres{0.015}{0.002}
& \gfpbres{0.0000}{0.0000} \\

& Velox
& \gres{0.004}{0.001}
& \gaucres{0.666}{0.023}
& \gres{0.000}{0.000}
& \gres{0.000}{0.000}
& \gres{0.000}{0.000}
& \gres{0.008}{0.003}
& \gfpbres{0.0000}{0.0000} \\

& Magic
& \gres{0.003}{0.001}
& \gaucbres{0.943}{0.004}
& \gres{0.000}{0.000}
& \gres{0.000}{0.000}
& \gres{0.000}{0.000}
& \gres{0.005}{0.002}
& \gfpbres{0.0000}{0.0000} \\

& \textbf{Theseus}
& \gres{0.002}{0.000}
& \gaucres{0.898}{0.015}
& \gres{0.000}{0.000}
& \gres{0.000}{0.000}
& \gres{0.000}{0.000}
& \gres{0.006}{0.003}
& \gfpres{0.0001}{0.0002} \\

\midrule

\multirow{5}{*}{\shortstack[c]{Trace\\(2.95)}}
& Allowlist
& ---
& ---
& \gres{0.057}{0.000}
& \gbres{0.107}{0.000}
& \gbres{0.216}{0.000}
& ---
& \gfpres{0.0003}{0.0000} \\

& Orthrus
& \gres{0.004}{0.002}
& \gaucres{0.888}{0.002}
& \gres{0.000}{0.000}
& \gres{0.000}{0.000}
& \gres{0.000}{0.000}
& \gres{0.020}{0.018}
& \gfpbres{0.0000}{0.0000} \\

& Velox
& \gbres{0.062}{0.005}
& \gaucres{0.868}{0.014}
& \gres{0.000}{0.000}
& \gres{0.000}{0.000}
& \gres{0.000}{0.001}
& \gres{0.105}{0.005}
& \gfpres{0.0008}{0.0011} \\

& Magic
& \gres{0.006}{0.001}
& \gaucbres{0.997}{0.000}
& \gres{0.000}{0.000}
& \gres{0.000}{0.000}
& \gres{0.000}{0.000}
& \gres{0.009}{0.001}
& \gfpbres{0.0000}{0.0000} \\

& \textbf{Theseus}
& \gres{0.026}{0.013}
& \gaucres{0.961}{0.005}
& \gbres{0.067}{0.149}
& \gres{0.008}{0.017}
& \gres{0.016}{0.037}
& \gbres{0.272}{0.246}
& \gfpbres{0.0000}{0.0001} \\

\midrule

\multirow{5}{*}{\shortstack[c]{Theia\\(6.72)}}
& Allowlist
& ---
& ---
& \gres{0.177}{0.000}
& \gbres{0.273}{0.000}
& \gbres{0.324}{0.000}
& ---
& \gfpres{0.0024}{0.0000} \\

& Orthrus
& \gres{0.003}{0.001}
& \gaucres{0.526}{0.095}
& \gres{0.000}{0.000}
& \gres{0.000}{0.000}
& \gres{-0.001}{0.000}
& \gres{0.010}{0.005}
& \gfpres{0.0004}{0.0001} \\

& Velox
& \gres{0.028}{0.011}
& \gaucres{0.491}{0.003}
& \gres{0.141}{0.005}
& \gres{0.097}{0.001}
& \gres{0.102}{0.002}
& \gbres{0.512}{0.029}
& \gfpres{0.0004}{0.0000} \\

& Magic
& \gres{0.088}{0.052}
& \gaucres{0.976}{0.004}
& \gres{0.000}{0.000}
& \gres{0.000}{0.000}
& \gres{0.000}{0.000}
& \gres{0.112}{0.043}
& \gfpbres{0.0000}{0.0000} \\

& \textbf{Theseus}
& \gbres{0.551}{0.150}
& \gaucbres{0.994}{0.002}
& \gbres{0.605}{0.305}
& \gres{0.251}{0.186}
& \gres{0.306}{0.198}
& \gres{0.469}{0.038}
& \gfpres{0.0001}{0.0001} \\

\bottomrule
\end{tabular}
\end{table*}

Theseus is implemented in PyTorch~\cite{pytorch} and PyTorch Geometric~\cite{pyg}. On E3, Theseus is trained with AdamW~\cite{loshchilov2018decoupled} for up to 400 epochs, with checkpoints selected using validation AP. Since ATLASv2 lacks an independent validation stage, Theseus and Velox use configurations fixed across h1 and h2. Theseus is trained for 300 epochs and Velox for 12 epochs, with the final checkpoint retained in each case. Thresholds are then calibrated using benign training scores, with ATLASv2 attack labels excluded from stopping decisions, checkpoint selection, and threshold calibration. Appendix~\ref{app:runtime} reports end-to-end runtime, graph construction cost, and memory usage for Theseus. We publicly release the code needed to reproduce the experiments.

\subsection{Metrics}
We report only operating-point metrics for the allowlist, whose binary output offers no ranking within the flagged and unflagged groups and no threshold sweep comparable to those of the learned models. For learned models, AP and AUROC summarize ranking quality across thresholds, while ADP summarizes attack coverage across thresholds. At the selected operating point, precision and FPR characterize alerting behavior, while F1 and MCC measure process-level recovery. Learned model results are reported as mean $\pm$ standard deviation over five random seeds.

\section{Results and Benchmark Analysis}\label{sec:experimental_results}

Table~\ref{tab:results_table} reports the main E3 results under the process-level scope. Cadets, FiveDirections, and Trace provide limited evidence for architectural comparison, whereas Theia is the only primary dataset on which Theseus clearly improves both anomaly ranking and node-level recovery over the other learned systems. These gains coincide with Theia's more complete and varied semantic fields, although differences in graph structure, workloads, attacks, and background activity prevent us from isolating the contribution of semantic signal.

\subsection{Primary E3 Results}

\subsubsection{Limited separation on Cadets, FiveDirections, and Trace}
On Cadets and Trace, some models surface attack scenarios, as reflected by precision at the selected threshold and by ADP across thresholds, but recover only a small portion of the labeled attack processes. FiveDirections provides little usable signal near the top of the ranking or at the selected operating point. The weak process-level recovery on these three datasets provides limited evidence for distinguishing architectures, since lexical novelty and structural heuristics remain plausible explanations for much of the observed detector behavior.

On Cadets, alerting and node-level recovery diverge sharply. Velox reaches high precision (0.883) with near-zero FPR, but its F1 is only 0.007. Its high ADP and low MCC reinforce this distinction, and Orthrus follows a similar pattern. On FiveDirections, all learned systems have near-zero AP and negligible recovery at the selected thresholds despite AUROC values ranging from 0.666 to 0.943. On Trace, AUROC ranges from 0.868 to 0.997, yet AP remains at or below 0.062 and fixed-threshold recovery is negligible. These AUROC values indicate that attack processes often score above benign processes, but the low AP shows that they are not sufficiently concentrated near the top of the ranking.

\begin{figure}[t]
    \centering
    \includegraphics[width=\columnwidth]{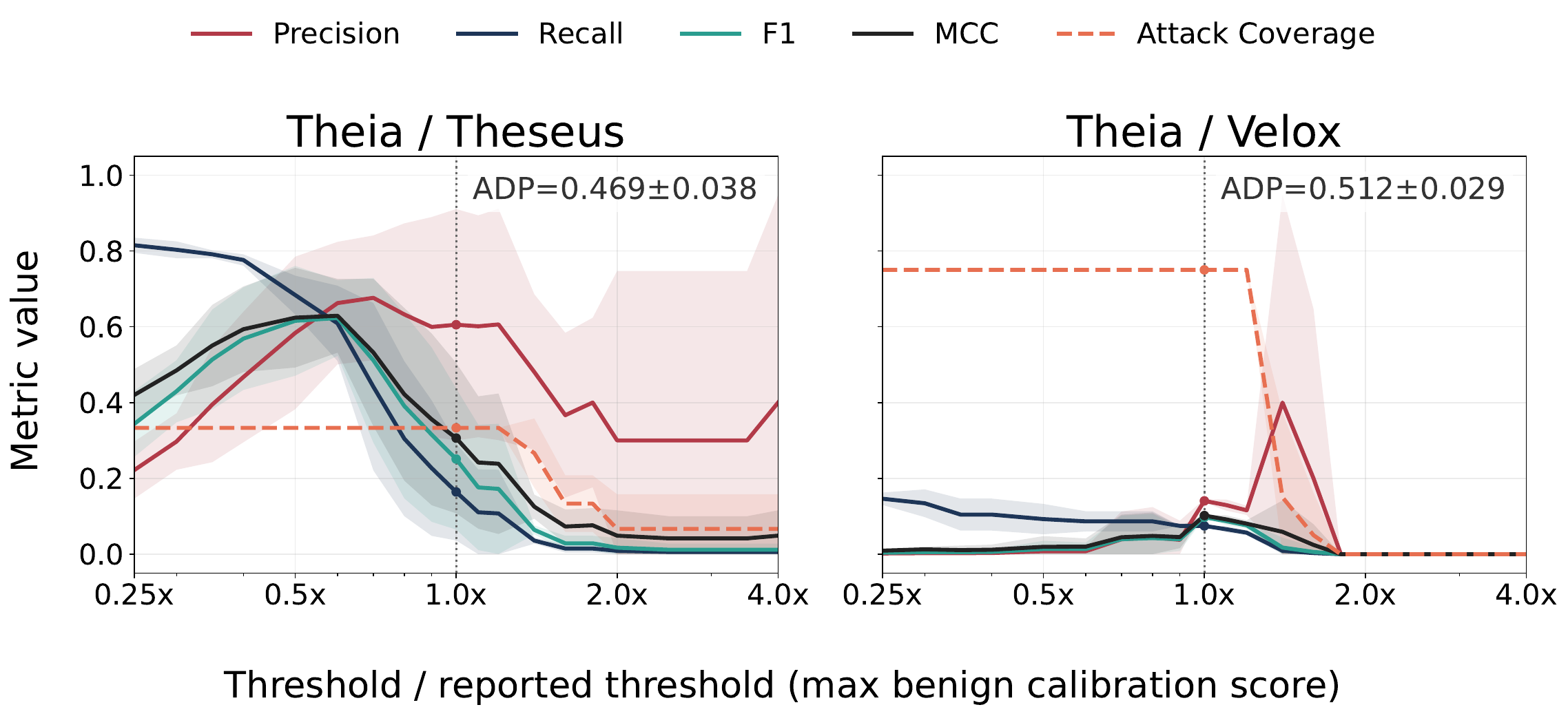}
    \caption{Threshold sensitivity on Theia for frozen Theseus and Velox models. Curves and annotated ADP values are means over five seeds, and shaded bands show one standard deviation. The horizontal axis shows the threshold relative to the reported threshold, set to the maximum benign validation score.}\label{fig:theia_threshold_sensitivity}
\end{figure}

\subsubsection{Lexical novelty diagnostic}
Using lexical novelty alone, the allowlist achieves the best F1 and MCC on all four datasets at the selected thresholds, as well as the best precision on FiveDirections. On Theia, Theseus has higher precision but slightly lower F1 and MCC than the allowlist. Appendix~\ref{app:command_allowlist} shows that exact command-line novelty generally raises recall but introduces many more false positives. These results show that strong operating-point metrics alone do not establish the benefit of learned behavioral modeling.

\subsubsection{Theia supports a clearer architectural comparison}
Theseus achieves the highest AP, AUROC, and precision on Theia. Its F1 and MCC exceed those of the other learned systems and approach those of the allowlist, with a lower FPR (0.0001 versus 0.0024). Its higher AP indicates recovery potential that the conservative operating point does not fully realize, and the threshold sweep in Section~\ref{sec:theia_threshold} shows that lower thresholds allow it to surpass the allowlist in F1 and MCC while maintaining a lower FPR. Magic reaches an AUROC of 0.976, but its AP is 0.088 and its thresholded recovery is zero, illustrating why AUROC must be interpreted alongside AP and operating-point metrics. Velox has higher ADP than Theseus, but lower AP, F1, and MCC. These differences make Theia the clearest setting for distinguishing the evaluated learned systems, consistent with its richer semantic information.

\subsubsection{Threshold sensitivity on Theia}\label{sec:theia_threshold}
To examine whether the recovery gap on Theia persists across thresholds, we evaluate frozen Theseus and Velox models across a range of decision thresholds, with the complete E3 curves reported in Appendix~\ref{app:threshold_sensitivity}. We choose Velox for this comparison because it achieves the highest F1 and MCC among the other learned systems on Theia at the reported threshold. Fig.~\ref{fig:theia_threshold_sensitivity} shows that lowering the Theseus threshold to $0.6{\times}$ the reported value increases its mean F1/MCC from 0.251/0.306 to 0.623/0.629, with an FPR of 0.0003. Both the $0.5{\times}$ and $0.6{\times}$ settings exceed the allowlist's F1 and MCC while retaining a lower FPR. Velox does not reach comparable recovery over the tested range. On Cadets and FiveDirections, recovery remains limited. On Trace, lowering the Velox threshold to $0.3{\times}$ yields F1/MCC of 0.162/0.241 with an FPR of 0.0026, demonstrating a meaningful change in the precision--recall trade-off. Changing the threshold substantially affects recovery on some datasets, but Velox does not reach the recovery achieved by Theseus on Theia within the tested range.

\begin{table}[t]
    \centering
    \small
    \caption{Semantic signal quality by dataset, computed as feature completeness multiplied by field entropy.}\label{tab:ssq_by_dataset}
    \setlength{\tabcolsep}{4pt}
    \begin{tabular}{l
                S[table-format=2.1]
                S[table-format=2.1]
                S[table-format=1.2]
                S[table-format=1.2]}
    \toprule
    \textbf{Dataset} &
    {\makecell[c]{\textbf{Nodes}\\\textbf{(M)}}} &
    {\makecell[c]{\textbf{Feature}\\\textbf{compl. (\%)}}} &
    {\makecell[c]{\textbf{Field entropy}\\\textbf{(bits)}}} &
    {\makecell[c]{\textbf{Semantic}\\\textbf{signal quality}}} \\
    \midrule
    Cadets   & 2.7  & 24.0 & 7.25 & 1.74 \\
    FiveDir. & 1.4  & 42.4 & 9.35 & 3.97 \\
    Trace    & 44.5 & 53.0 & 5.56 & 2.95 \\
    Theia    & 1.5  & 87.0 & 7.72 & 6.72 \\
    \bottomrule
\end{tabular}
\end{table}

\begin{figure}[t]
    \centering
    \includegraphics[trim={2.75em 1em 0 0},width=0.67\linewidth]{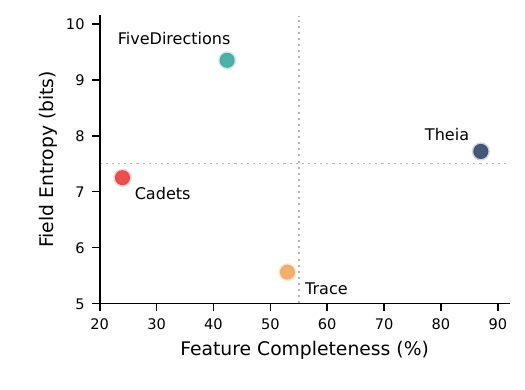}
    \caption{Feature completeness and field entropy in the evaluated E3 datasets. Theia has the strongest combination of populated semantic attributes and token diversity, while Cadets, FiveDirections, and Trace are constrained by different forms of missing or shallow signal.}\label{fig:semantic_signal_scatter}
\end{figure}

\subsection{Semantic Signal Quality and Dataset Separation}
\label{sec:ssq_analysis}

The results on Theia motivate examining whether architectural separation is related to the semantic evidence exposed by each dataset. We use the semantic signal quality diagnostic defined in Section~\ref{sec:semantic_signal}, which combines feature completeness and field entropy in the processed node tables. Fig.~\ref{fig:semantic_signal_scatter} visualizes this dataset-level pattern, and Table~\ref{tab:ssq_by_dataset} reports the corresponding values. The node counts in that table refer to rows in the processed node tables, while Table~\ref{tab:dataset_stats} counts node instances across temporal snapshots. Because files and network flows contribute to graph representation and message passing even though metrics are computed only over processes, we compute $Q$ across all three node types. Table~\ref{tab:ssq_breakdown} reports the decomposition by node type.

Across the audited E3 datasets, the ordering of $Q$ is consistent with the observed architectural separation: Theseus has no consistent advantage over the learned baselines on Cadets, Trace, or FiveDirections, but shows clear gains in AP, F1, and MCC on Theia. We use $Q$ as a descriptive diagnostic of the attributes exposed by our preprocessing and tokenization to interpret these results, without assuming that it predicts performance on other benchmarks.

\begin{table}[t]
    \centering
    \small
    \caption{Feature completeness and average field entropy by node type.}\label{tab:ssq_breakdown}
    \begin{tabular}{ll
                S[table-format=3.1]
                S[table-format=2.2]}
        \toprule
        \textbf{Dataset} &
        \textbf{Node type} &
        {\makecell[c]{\textbf{Feature}\\\textbf{compl. (\%)}}} &
        {\makecell[c]{\textbf{Avg. field}\\\textbf{entropy (bits)}}} \\
        \midrule
        \multirow{3}{*}{Cadets} & Process & 49.9  & 1.91 \\
         & File    & 13.9  & 10.29 \\
         & Network & 100.0 & 5.94 \\
        \midrule
        \multirow{3}{*}{FiveDirections} & Process & 1.5   & 4.06 \\
         & File    & 95.6  & 9.65 \\
         & Network & 70.7  & 7.19 \\
        \midrule
        \multirow{3}{*}{Trace} & Process & 50.0  & 2.75 \\
         & File    & 98.1  & 9.94 \\
         & Network & 100.0 & 4.96 \\
        \midrule
        \multirow{3}{*}{Theia} & Process & 99.6  & 5.06 \\
         & File    & 77.7  & 9.67 \\
         & Network & 100.0 & 7.39 \\
        \bottomrule
    \end{tabular}
\end{table}

\begin{figure*}[!t]
     \centering
     \subfloat[Theseus\label{fig:theseus_cadets}]{
        \includegraphics[height=0.118\textwidth]{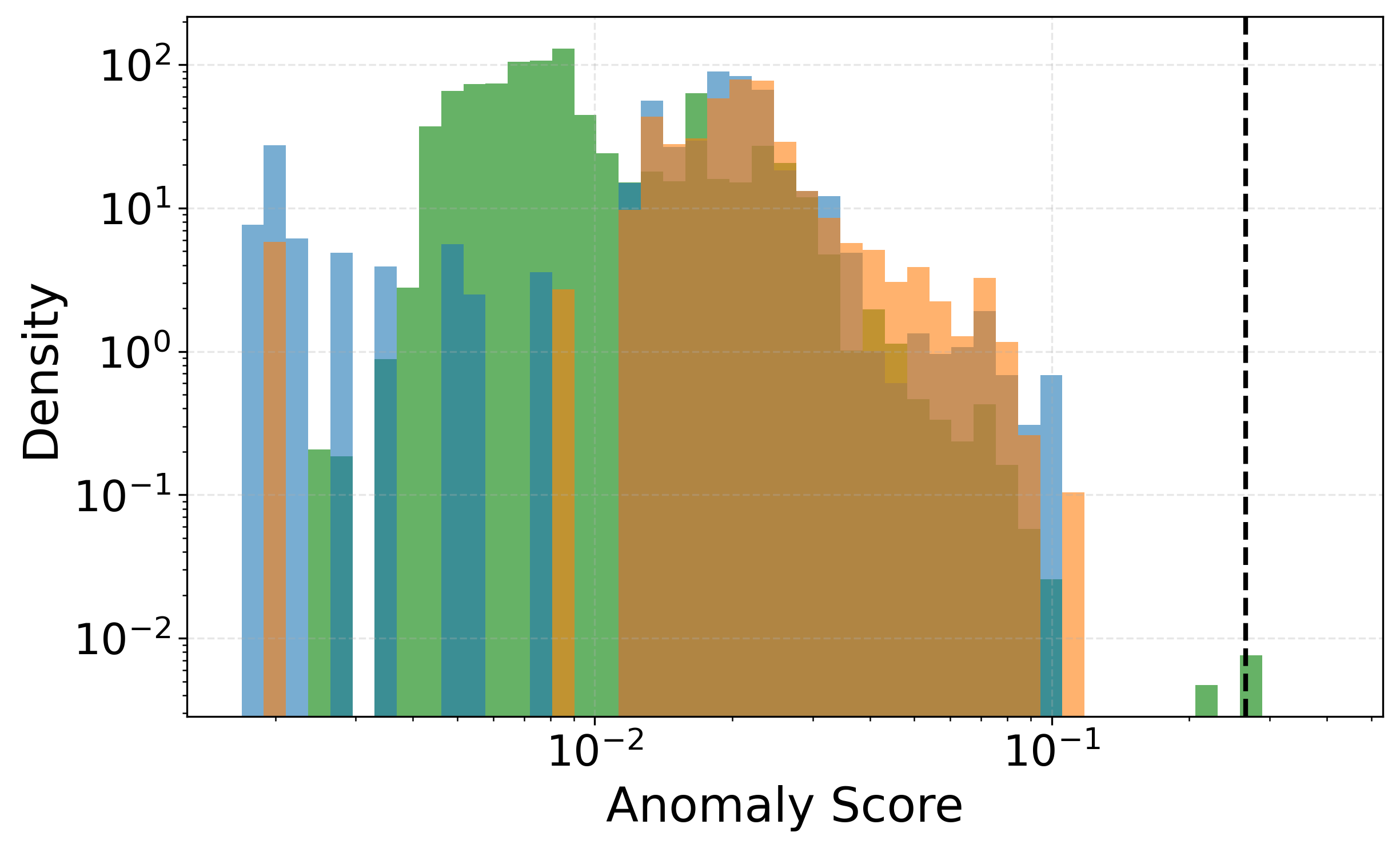}
     }
     \hfill
     \subfloat[Magic\label{fig:magic_cadets}]{
        \includegraphics[height=0.118\textwidth]{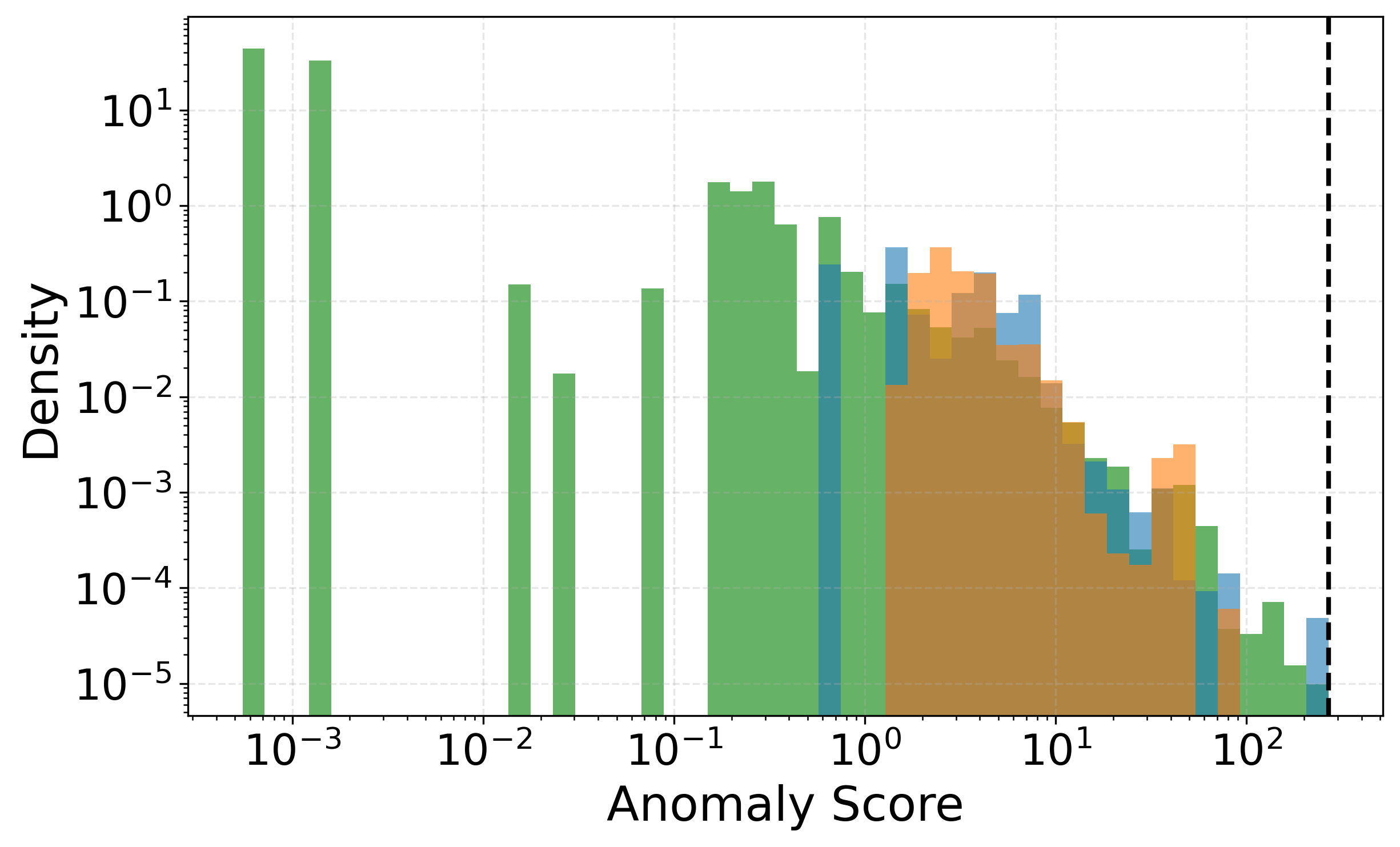}
     }
     \hfill
     \subfloat[Velox\label{fig:velox_cadets}]{
        \includegraphics[height=0.118\textwidth]{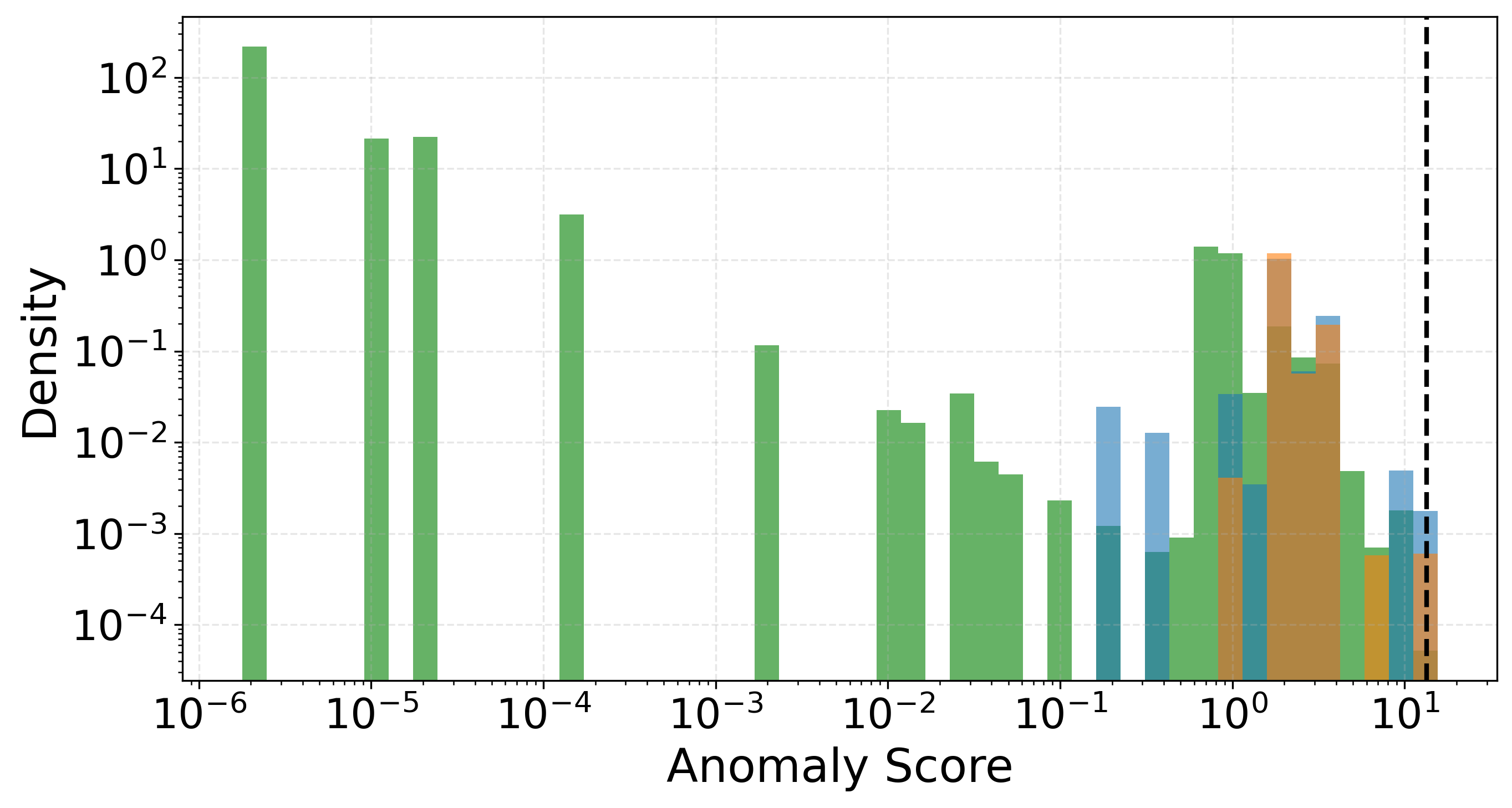}
     }
     \hfill
     \subfloat[Orthrus\label{fig:orthrus_cadets}]{
        \includegraphics[height=0.118\textwidth]{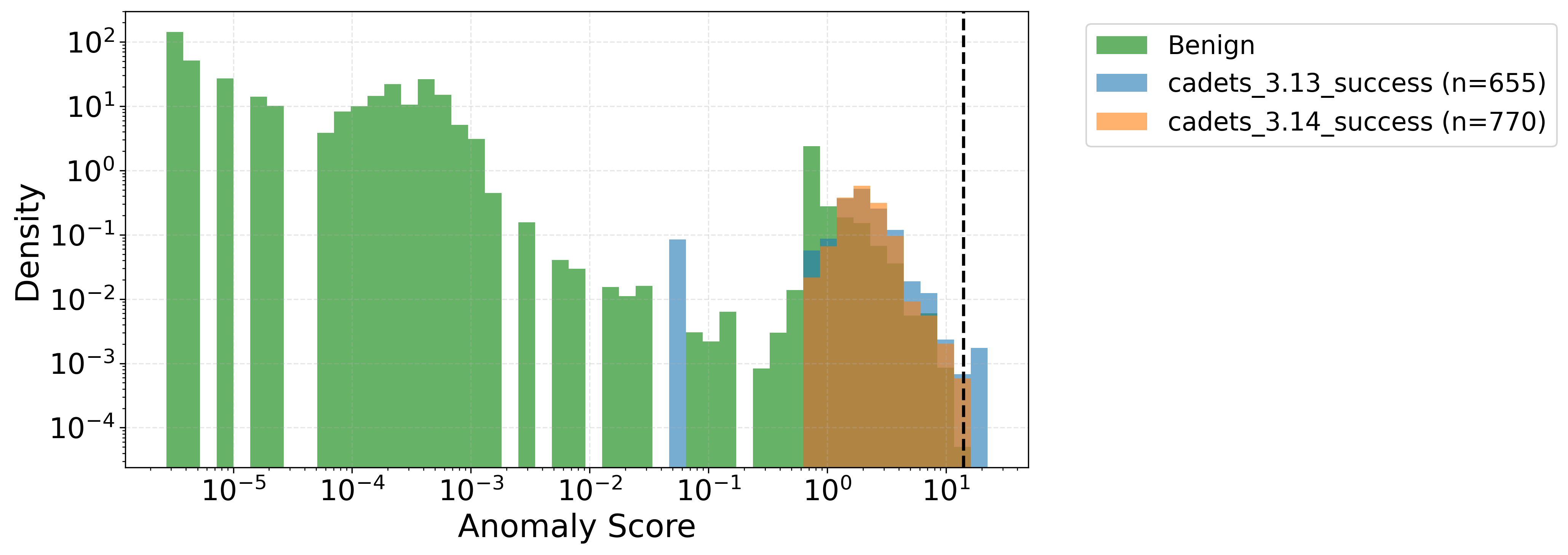}
     }
      \caption{Anomaly score distributions by attack type for Cadets.}\label{fig:cadets_anomaly_scores}
      \Description{Four subplots comparing anomaly score distributions on Cadets for Theseus, Magic, Velox, and Orthrus. All models show substantial overlap between benign and malicious scores, with weaker separation than on Theia.}
\end{figure*}

\begin{figure*}[!t]
     \centering
     \subfloat[Theseus\label{fig:theseus_fivedirections}]{
        \includegraphics[height=0.118\textwidth]{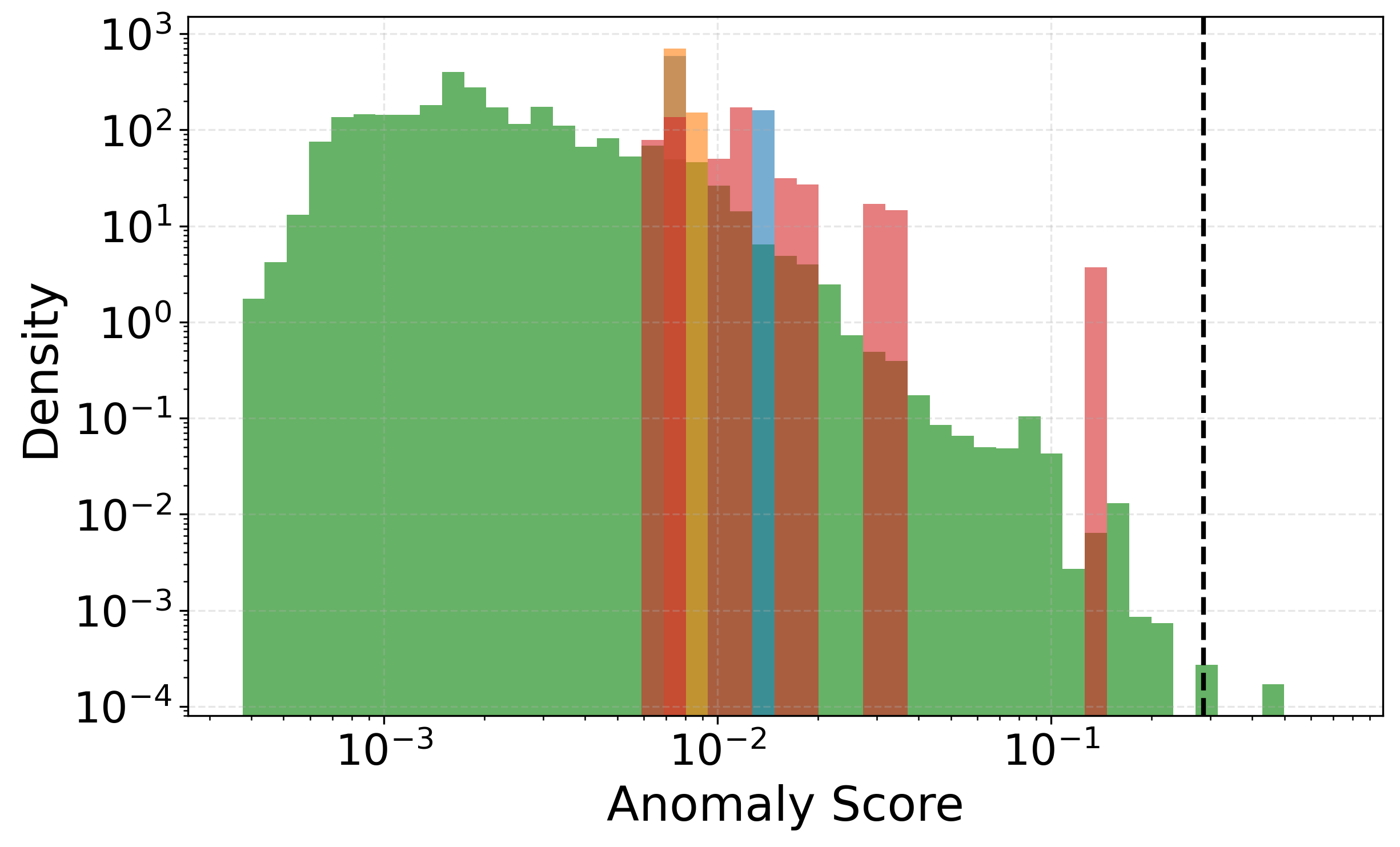}
     }
     \hfill
     \subfloat[Magic\label{fig:magic_fivedirections}]{
        \includegraphics[height=0.118\textwidth]{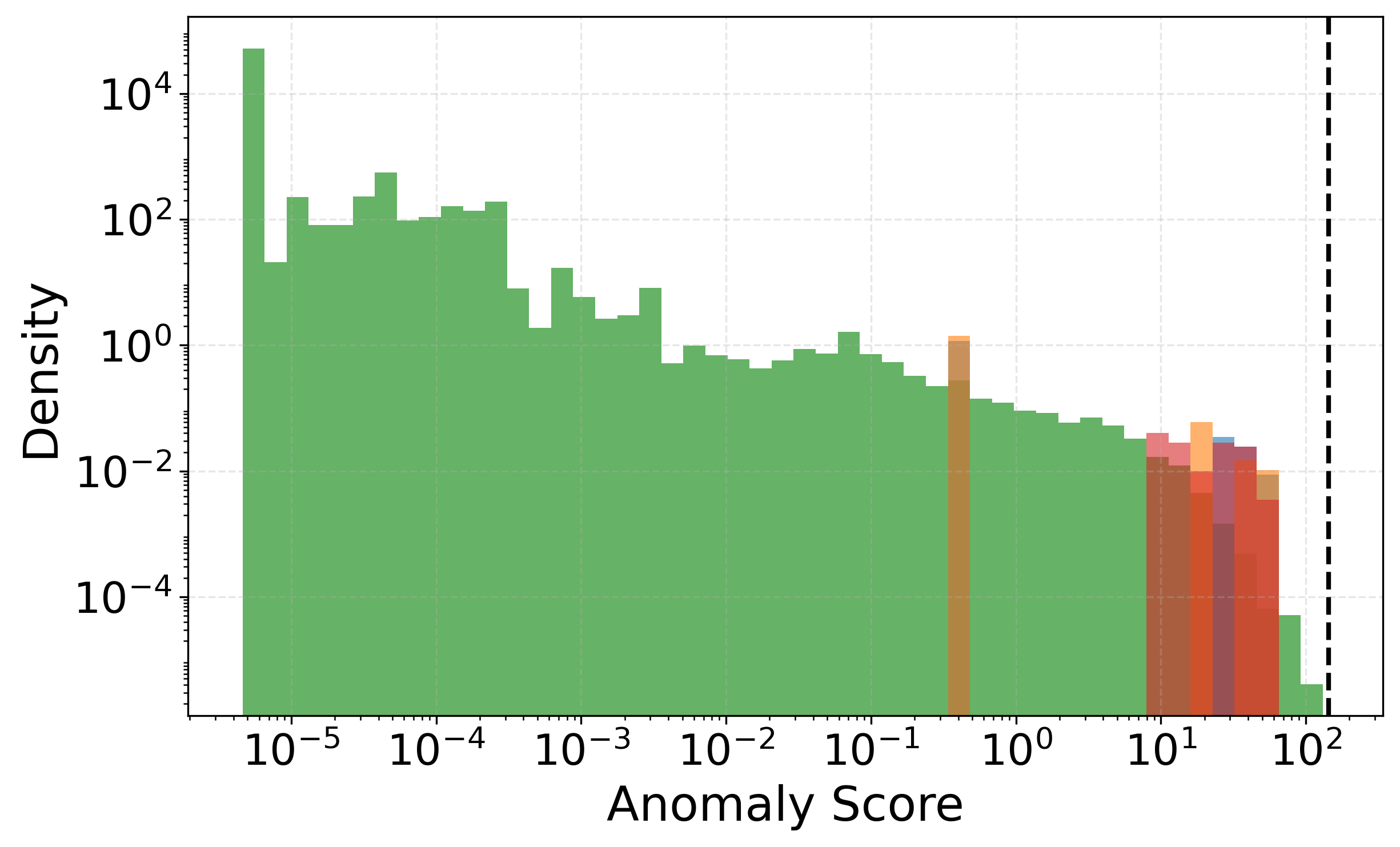}
     }
     \hfill
     \subfloat[Velox\label{fig:velox_fivedirections}]{
        \includegraphics[height=0.118\textwidth]{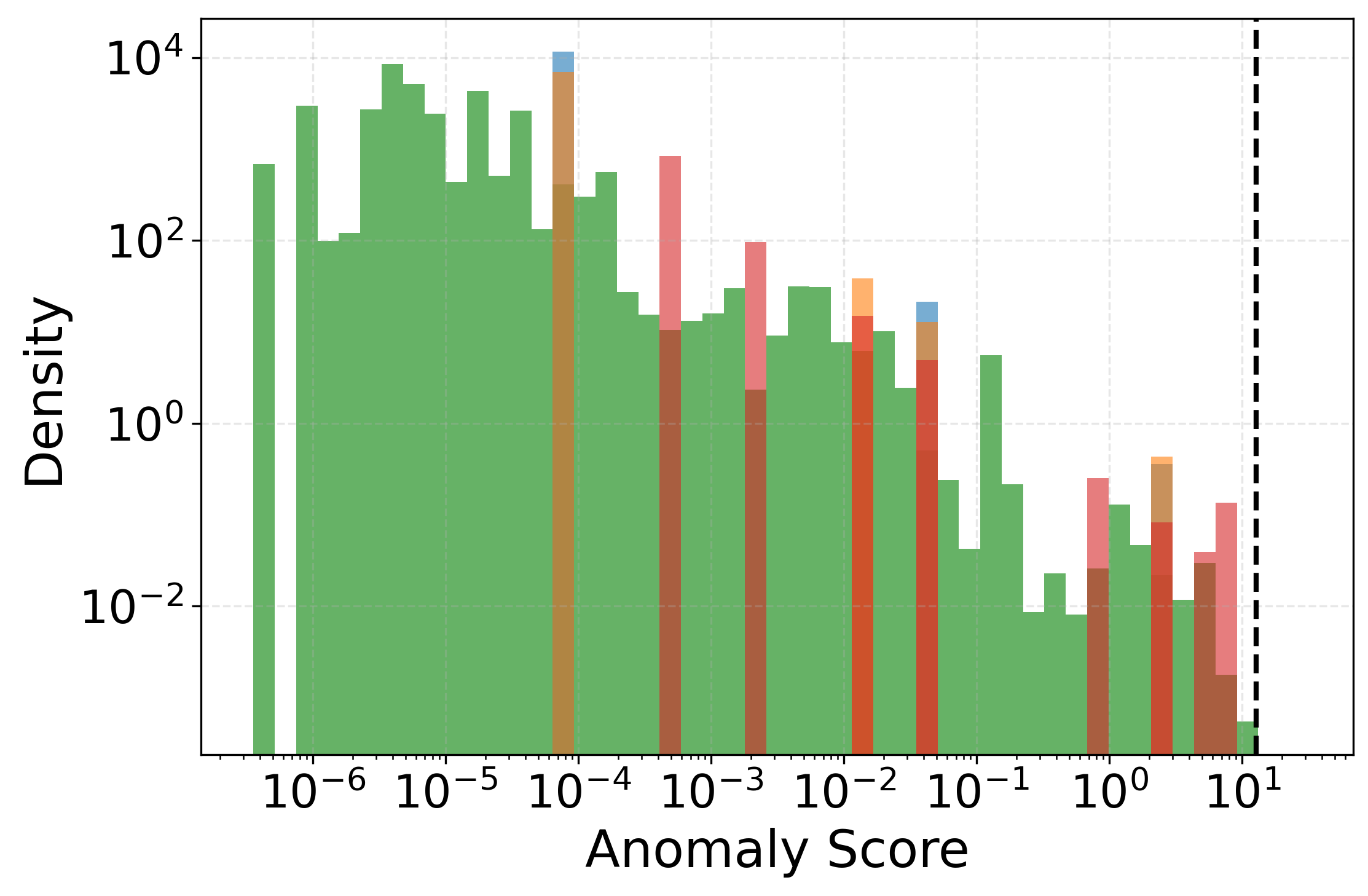}
     }
     \hfill
     \subfloat[Orthrus\label{fig:orthrus_fivedirections}]{
        \includegraphics[height=0.118\textwidth]{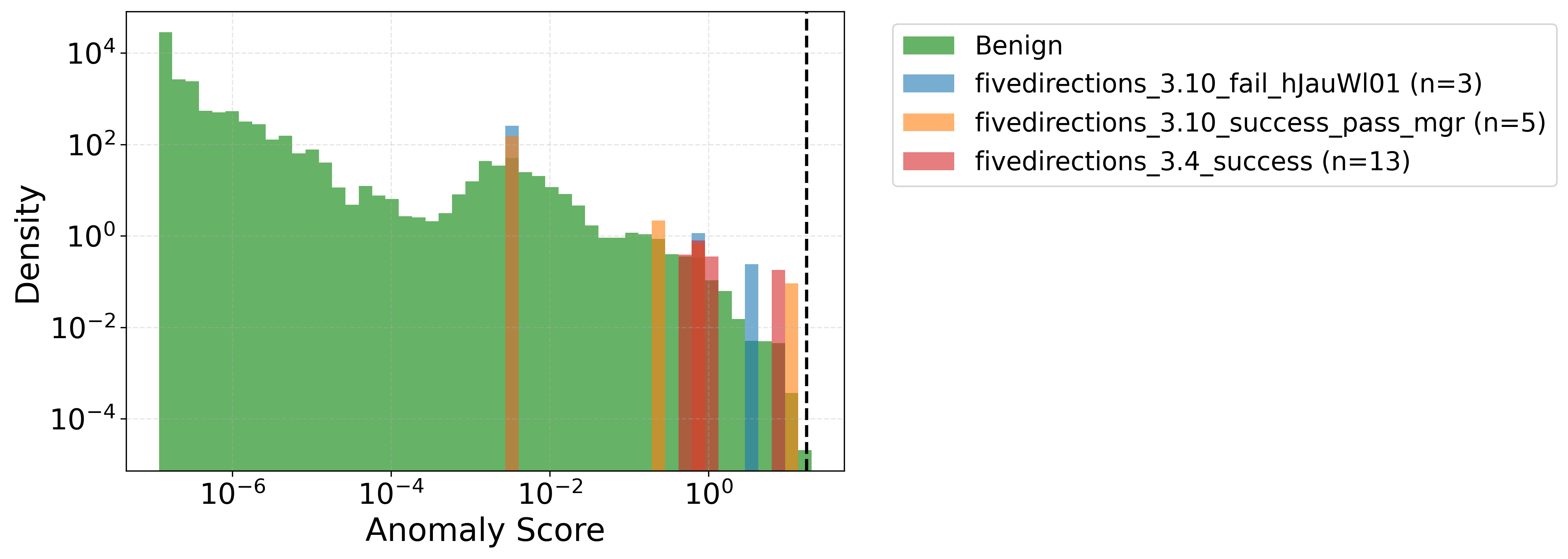}
     }
      \caption{Anomaly score distributions by attack type for FiveDirections.}\label{fig:fivedirections_anomaly_scores}
      \Description{Four subplots comparing anomaly score distributions on FiveDirections for Theseus, Magic, Velox, and Orthrus. The score distributions are sparse and only weakly separated, reflecting limited usable signal.}
\end{figure*}

Among the primary E3 datasets, Theia is the only one pairing high completeness (87.0\%) with high entropy (7.72 bits); Cadets, Trace, and FiveDirections are constrained by sparsity, repetition, or inconsistent coverage.

Trace contains 44.5 million records in the processed node tables, but its completeness is moderate (53.0\%) and its entropy is the lowest in the suite (5.56 bits).

Cadets is constrained by both sparsity and shallow process semantics. Although its populated fields have an average entropy of 7.25 bits, the dataset reaches only 24.0\% completeness. Its process command vocabulary is also highly restricted, with 130 unique strings across nearly 225,000 processes and 66 tokens under the Theseus tokenizer. Our analysis of the source data shows that the first 155,000 processes already cover 90\% of this vocabulary. A few simple commands, such as \texttt{sleep} and \texttt{top}, dominate the distribution, leaving little variation in command semantics for a model to learn.

FiveDirections combines diverse populated attributes with sparse coverage. Its populated attributes have the highest entropy in E3 (9.35 bits), but only 42.4\% of applicable attribute slots are populated. The raw DARPA logs contain the largest vocabulary of command tokens in the suite, yet commands are populated for only 3.1\% of processes. Since most processes lack a command line, more expressive models have little semantic information to exploit on this dataset.

Higher entropy indicates greater token diversity, but that variation may include noise or incidental details with little value for detection. Together with completeness, it helps distinguish datasets with sparse or repetitive attributes from those exposing more varied information for models to learn from. These patterns should also be read alongside Liu et al.'s audit~\cite{liu2025what}, as the values reflect how the data were collected and represented. FiveDirections' high entropy may partly reflect Windows-specific vocabulary, while Trace's low entropy and completeness may partly reflect execution partitioning, with execution-unit nodes accounting for a large share of the analyzed entities.

\begin{figure*}
     \centering
     \subfloat[Theseus\label{fig:theseus_trace}]{
        \includegraphics[height=0.118\textwidth]{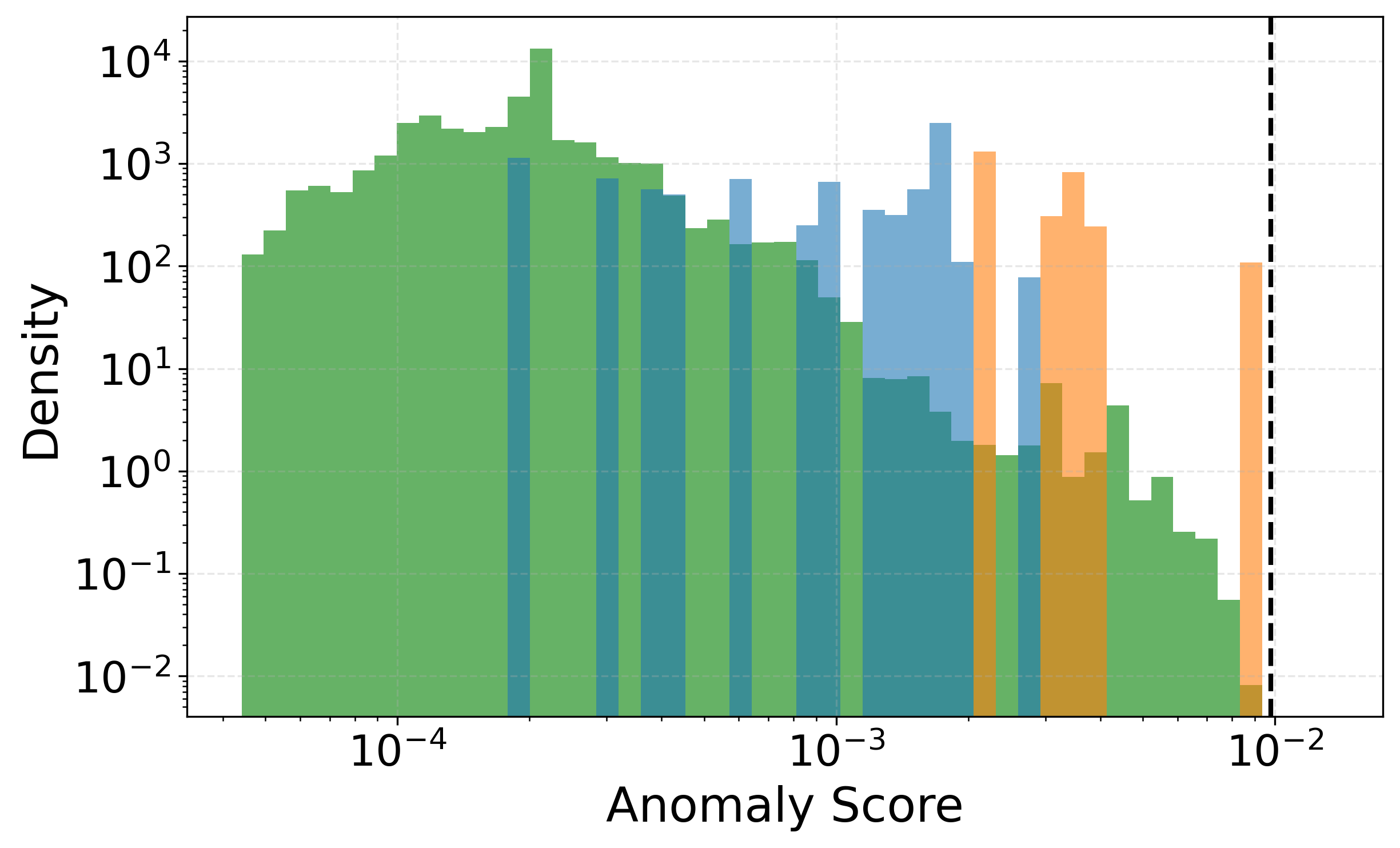}
     }
     \hfill
     \subfloat[Magic\label{fig:magic_trace}]{
        \includegraphics[height=0.118\textwidth]{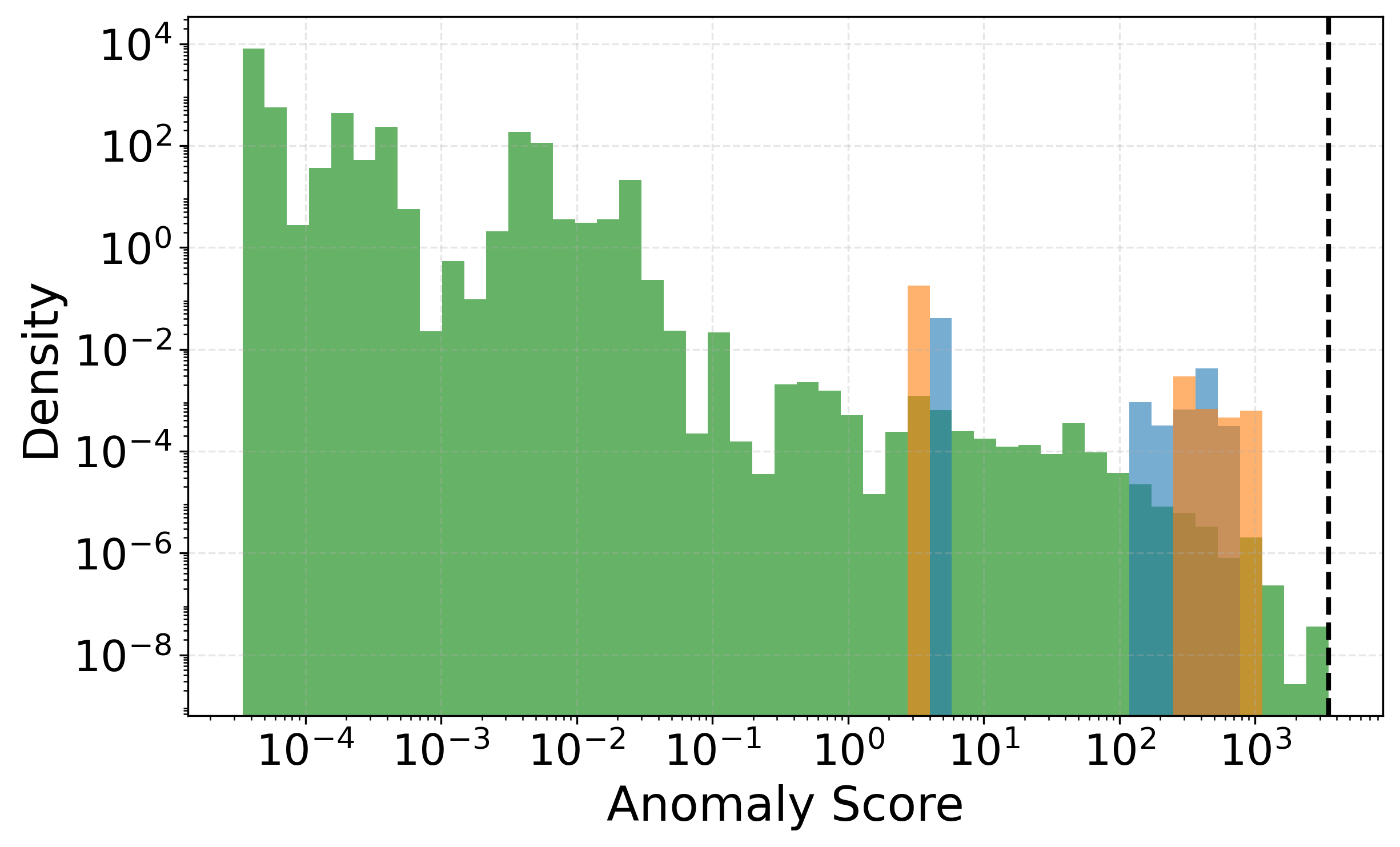}
     }
     \hfill
     \subfloat[Velox\label{fig:velox_trace}]{
        \includegraphics[height=0.118\textwidth]{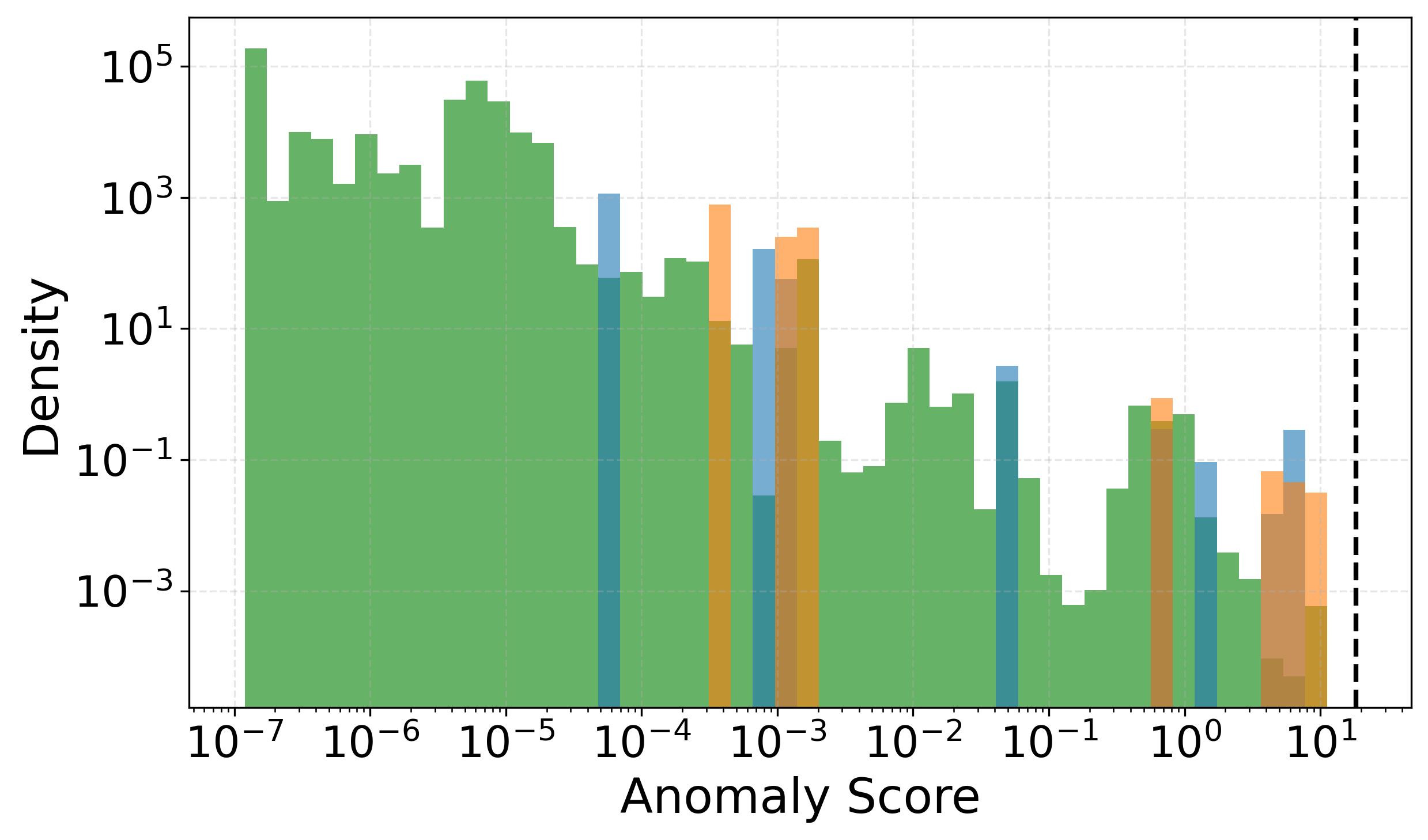}
     }
     \hfill
     \subfloat[Orthrus\label{fig:orthrus_trace}]{
        \includegraphics[height=0.118\textwidth]{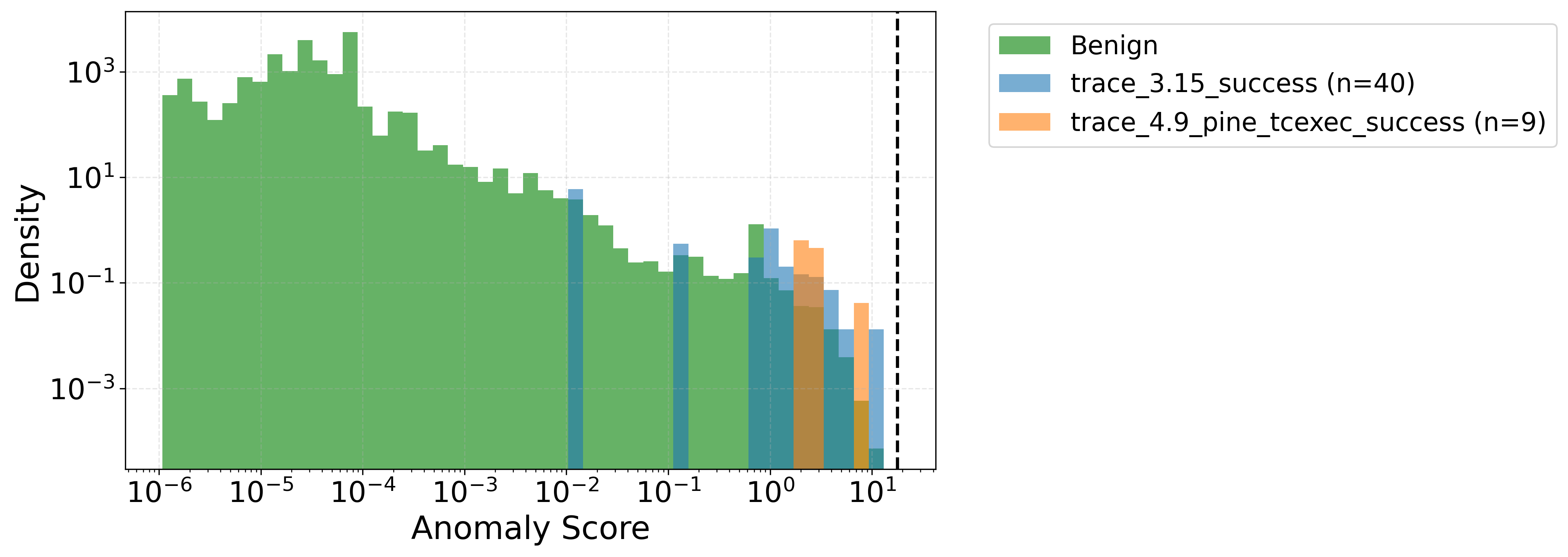}
     }
      \caption{Anomaly score distributions by attack type for Trace.}\label{fig:trace_anomaly_scores}
      \Description{Four subplots comparing anomaly score distributions on Trace for Theseus, Magic, Velox, and Orthrus. Most malicious and benign scores remain overlapped, consistent with weak semantic differentiation on this dataset.}
\end{figure*}

\begin{figure*}
     \centering
     \subfloat[Theseus\label{fig:theseus_theia}]{
        \includegraphics[height=0.118\textwidth]{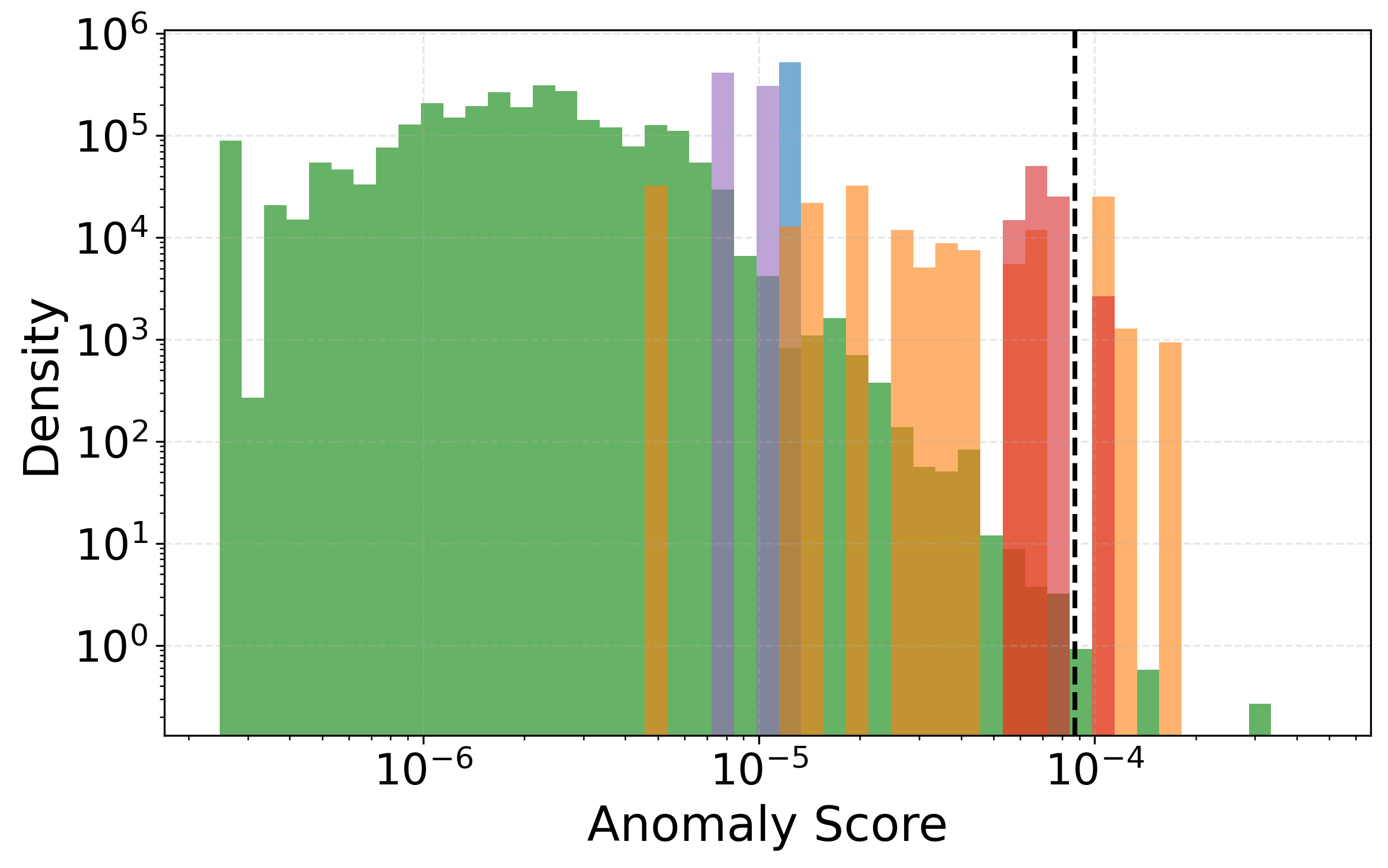}
     }
     \hfill
     \subfloat[Magic\label{fig:magic_theia}]{
        \includegraphics[height=0.118\textwidth]{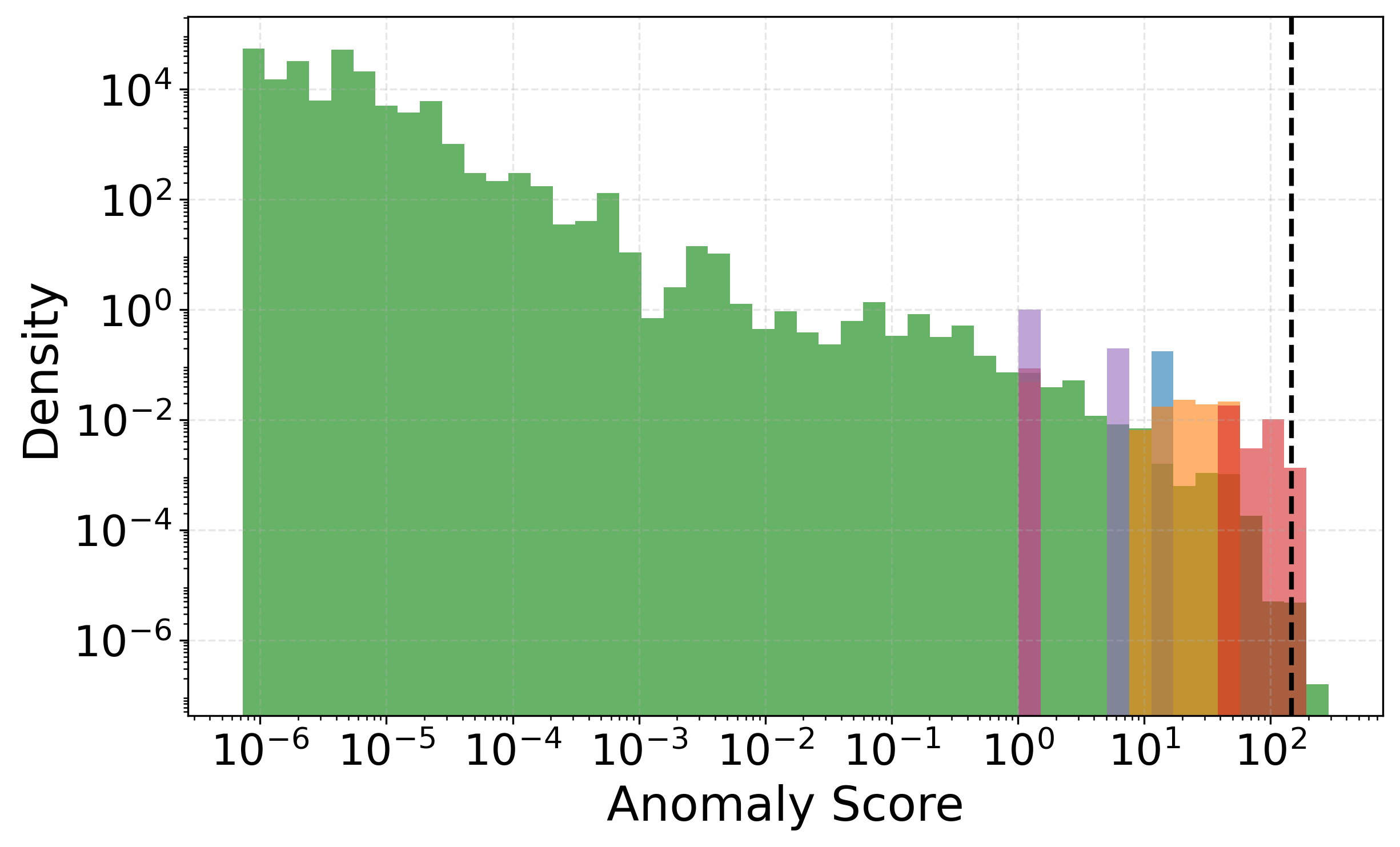}
     }
     \hfill
     \subfloat[Velox\label{fig:velox_theia}]{
        \includegraphics[height=0.118\textwidth]{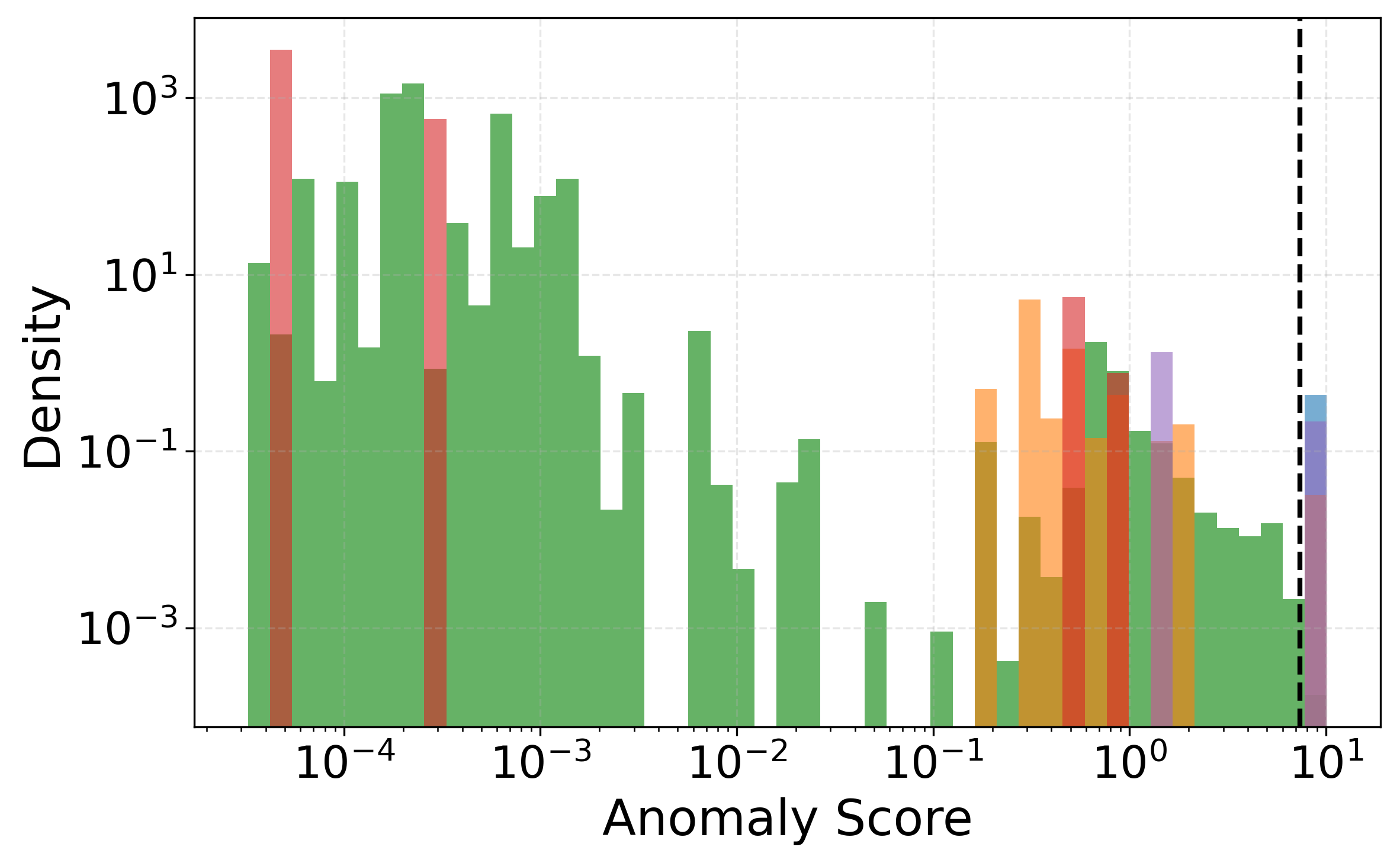}
     }
     \hfill
     \subfloat[Orthrus\label{fig:orthrus_theia}]{
        \includegraphics[height=0.118\textwidth]{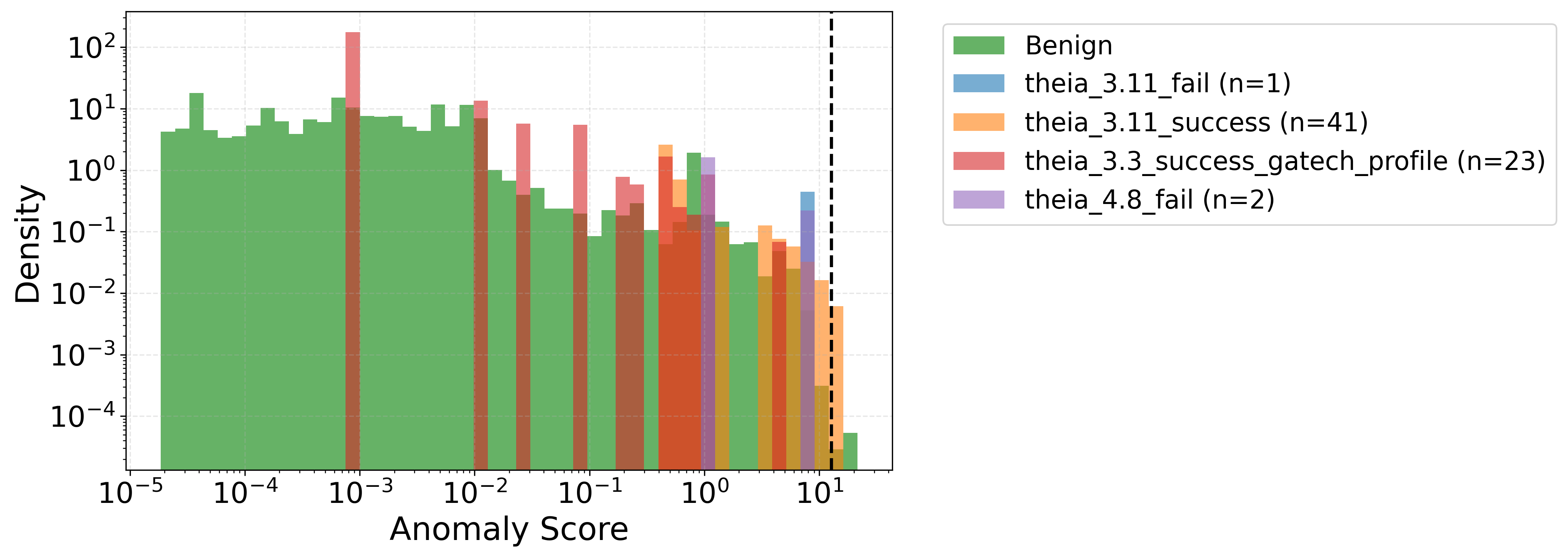}
     }
      \caption{Anomaly score distributions by attack type for Theia.}\label{fig:theia_anomaly_scores}
      \Description{Four subplots comparing anomaly score distributions on Theia for Theseus, Magic, Velox, and Orthrus. Theseus shows the clearest separation between benign and malicious score ranges, while the baselines exhibit heavier overlap.}
\end{figure*}

\subsection{Anomaly Score Distributions}

Figs.~\ref{fig:cadets_anomaly_scores}--\ref{fig:theia_anomaly_scores} show the score distributions underlying Table~\ref{tab:results_table}. On Theia, Theseus places many attack processes in the right tail, although scores for some attack scenarios overlap with benign scores. This is consistent with the main results, where Theseus achieves the strongest AP, F1, and MCC among the learned systems on Theia.

On FiveDirections and Trace, scores for malicious processes tend to be higher but overlap substantially with benign scores. This can yield high AUROC because many malicious--benign pairs are ordered correctly, while AP remains low because attack processes are rare and insufficiently concentrated near the top of the ranking. The same overlap leaves few attack processes above the conservative threshold, so recovery at the selected operating point is negligible.

On Cadets, scores for attack processes are broadly spread, with higher values overlapping those of benign processes. This overlap is consistent with novelty-based scoring, where unfamiliar attributes at test time raise anomaly scores for both attack nodes and unusual benign nodes. A model can consequently achieve high ADP while still producing poor node-level recovery, as reflected by low F1 and MCC.

\subsection{Theseus Sensitivity and Ablations}\label{sec:ablations}

We examine process creation events, token weighting, and structural features by varying one aspect of the Theseus reference configuration used in Table~\ref{tab:results_table} at a time, with the remaining settings held fixed. We report AP to measure ranking performance, together with operating-point metrics to show how each change affects recovery.

\subsubsection{Process creation events}\label{sec:process_creation}

\begin{table}[t]
\centering
\footnotesize
\setlength{\tabcolsep}{4pt}
\caption{Sensitivity to process creation events on Theia.}
\label{tab:process_creation}
\begin{tabular}{lrccc}
\toprule
\makecell[c]{\textbf{Creation}\\\textbf{events}}
& \makecell[c]{\textbf{Training}\\\textbf{entities}}
& \textbf{AP}
& \textbf{F1}
& \textbf{MCC} \\
\midrule
Excluded
& 244{,}605
& \bres{0.650}{0.075}
& \bres{0.381}{0.182}
& \bres{0.451}{0.174} \\
Included
& 265{,}446
& \res{0.551}{0.150}
& \res{0.251}{0.186}
& \res{0.306}{0.198} \\
\bottomrule
\end{tabular}
\end{table}

The E3 datasets record process creation under different event names. In the source event tables, Cadets and FiveDirections contain only \texttt{EVENT\_FORK}, Theia contains only \texttt{EVENT\_CLONE}, and Trace contains both. We map both names to a common encoding to preserve these events and represent process creation consistently across datasets. Table~\ref{tab:process_creation} examines sensitivity to this event selection on Theia by comparing graphs with and without process creation events. Exclusion increases mean AP from 0.551 to 0.650 and also improves F1 and MCC.

Training entities are identified from the endpoints of events kept during preprocessing, so including process creation events changes both graph context and entity coverage. On Theia, it adds edges and increases the training pool from 244,605 to 265,446 unique entities. The performance difference thus reflects changes in both the available interactions and the entities used for learning.

\subsubsection{Token weighting}
\begin{table}[t]
\centering
\footnotesize
\caption{Token weighting ablation on Theia under the shared threshold policy.}
\label{tab:weighting_ablation}
\setlength{\tabcolsep}{5pt}
\begin{tabular}{lccc}
\toprule
\textbf{Configuration} & \textbf{AP $\uparrow$} & \textbf{MCC $\uparrow$} & \textbf{FPR $\downarrow$} \\
\midrule
Standard decay & \res{0.422}{0.106} & \res{0.014}{0.032} & \bres{0.0000}{0.0000} \\
Reverse files & \res{0.313}{0.118} & \res{0.064}{0.091} & \res{0.0001}{0.0000} \\
Reverse network & \bres{0.551}{0.150} & \res{0.306}{0.198} & \res{0.0001}{0.0001} \\
Reverse both & \res{0.478}{0.118} & \bres{0.330}{0.241} & \res{0.0001}{0.0000} \\
\bottomrule
\end{tabular}
\end{table}

Table~\ref{tab:weighting_ablation} reports the token weighting ablation on Theia. Reverse network flow weighting achieves the highest mean AP and improves MCC over standard decay, suggesting that destination services are more informative than transient source ports. Reversing both file and network weights yields a slightly higher mean MCC, but lower AP.

\subsubsection{Structural features}

We evaluated edge counts and node degrees as explicit structural features. These features can help when semantic attributes are sparse, but they can also emphasize volume and degree patterns when semantic fields already provide stronger evidence.

Table~\ref{tab:features_ablation} shows this trade-off. Adding structural features reduces AP on Theia from 0.551 to 0.312, suggesting that graph density cues interfere with the stronger semantic signal present in that dataset. Structural features improve mean AP on Cadets, FiveDirections, and Trace, although process recovery remains limited. This supports the broader interpretation that models rely more on structural heuristics when semantic fields are incomplete or repetitive.

\begin{table}[t]
\centering
\footnotesize
\setlength{\tabcolsep}{5pt}
\caption{Structural feature ablation under the shared threshold policy. Bold values indicate the best mean, including ties.}\label{tab:features_ablation}
\begin{tabular}{ll ccc}
        \toprule
        \textbf{Dataset} & \textbf{Config} & \textbf{AP $\uparrow$} & \textbf{MCC $\uparrow$} & \textbf{ADP $\uparrow$} \\
    
    \midrule
        Cadets & W/O & \res{0.089}{0.023} & \bres{-0.001}{0.001} & \bres{0.434}{0.130} \\
         & W/ & \bres{0.215}{0.093} & \bres{-0.001}{0.002} & \res{0.400}{0.058} \\
        \midrule
        FiveDir. & W/O & \res{0.001}{0.000} & \bres{0.000}{0.000} & \res{0.001}{0.001} \\
         & W/ & \bres{0.002}{0.000} & \bres{0.000}{0.000} & \bres{0.006}{0.003} \\
        \midrule
        Trace & W/O & \res{0.015}{0.007} & \res{0.000}{0.000} & \res{0.023}{0.011} \\
         & W/ & \bres{0.026}{0.013} & \bres{0.016}{0.037} & \bres{0.272}{0.246} \\
        \midrule
        Theia & W/O & \bres{0.551}{0.150} & \bres{0.306}{0.198} & \bres{0.469}{0.038} \\
         & W/ & \res{0.312}{0.080} & \res{0.112}{0.030} & \res{0.335}{0.044} \\
        \bottomrule
\end{tabular}
\end{table}

On Cadets, structural features improve AP, while ADP decreases and MCC remains near zero. ADP should therefore be read alongside AP and MCC, since textual novelty on Cadets can support alerting without reliable attack reconstruction. Combined with the allowlist diagnostic, the structural feature ablation suggests that much of the measurable signal in these benchmarks comes from textual and attribute fields.

\subsection{Secondary ATLASv2 Results}
\begin{table*}[!t]
\centering
\small
\setlength{\tabcolsep}{3.3pt}
\caption{ATLASv2 secondary results using UUID-based REAPr process labels. Learned methods use benign training scores for threshold calibration because each host has a single attack window and no independent validation split containing attack data. \(Q\) denotes semantic signal quality. The allowlist is deterministic.}
\label{tab:atlas_results}
\begin{tabular}{cl ccccccc}
\toprule
\textbf{Host ($Q$)}
& \textbf{System}
& \textbf{AP $\uparrow$}
& \textbf{AUROC $\uparrow$}
& \textbf{Precision $\uparrow$}
& \textbf{F1 Score $\uparrow$}
& \textbf{MCC $\uparrow$}
& \textbf{ADP $\uparrow$}
& \textbf{FPR $\downarrow$} \\
\midrule

\multirow{3}{*}{\shortstack[c]{ATLASv2-h1\\(7.29)}}
& Allowlist
& ---
& ---
& \gbres{0.641}{0.000}
& \gbres{0.364}{0.000}
& \gbres{0.397}{0.000}
& ---
& \gfpres{0.0030}{0.0000} \\

& Velox
& \gres{0.193}{0.040}
& \gaucbres{0.918}{0.004}
& \res{0.000}{0.000}
& \res{0.000}{0.000}
& \res{0.000}{0.000}
& \gres{0.394}{0.184}
& \gfpbres{0.0000}{0.0000} \\

& \textbf{Theseus}
& \gbres{0.216}{0.026}
& \gaucres{0.911}{0.004}
& \gres{0.530}{0.313}
& \gres{0.173}{0.098}
& \gres{0.229}{0.129}
& \gbres{0.826}{0.122}
& \gfpres{0.0012}{0.0010} \\

\midrule

\multirow{3}{*}{\shortstack[c]{ATLASv2-h2\\(5.74)}}
& Allowlist
& ---
& ---
& \gbres{0.545}{0.000}
& \gbres{0.462}{0.000}
& \gbres{0.465}{0.000}
& ---
& \gfpres{0.0014}{0.0000} \\

& Velox
& \gbres{0.299}{0.015}
& \gaucbres{0.987}{0.000}
& \res{0.000}{0.000}
& \res{0.000}{0.000}
& \res{$-$0.001}{0.000}
& \gres{0.410}{0.047}
& \gfpbres{0.0002}{0.0001} \\

& \textbf{Theseus}
& \gres{0.267}{0.089}
& \gaucres{0.938}{0.007}
& \gres{0.411}{0.390}
& \gres{0.265}{0.258}
& \gres{0.283}{0.273}
& \gbres{0.698}{0.211}
& \gfpres{0.0005}{0.0004} \\

\bottomrule
\end{tabular}
\end{table*}

We evaluate ATLASv2 to check whether the main pattern extends beyond the DARPA CDM pipeline. For this evaluation, we use ATLASv2's Carbon Black telemetry to reconstruct a process-centric graph, which differs from E3 in schema and graph topology. As described in Section~\ref{sec:atlas_methodology}, the UUID-based REAPr process labels map directly to the reconstructed Carbon Black process nodes. Because each host contains a single attack period and the scenarios share substantial payload structure, thresholds are calibrated on benign training data only.

Table~\ref{tab:atlas_results} shows mixed architectural separation in this secondary setting. Theseus achieves higher AP on the higher-entropy h1 and higher ADP on both hosts, whereas Velox achieves higher AP on h2 and higher AUROC on both hosts. At the selected thresholds, Theseus achieves positive F1 and MCC while Velox has zero F1, although the allowlist achieves the strongest precision, F1, and MCC on both hosts. Theseus's AP advantage on the higher-entropy host is consistent with the E3 observations, but two hosts are insufficient to establish a general relationship. Given the graph reconstruction and split limitations described above, ATLASv2 provides corroborative evidence.

\section{Limitations}\label{sec:limitations}

\subsection{Scope and Generality}

Our comparative claims are limited to the audited DARPA TC E3 suite, with ATLASv2 used only as secondary evidence. The primary suite provides a consistent process-label target, public artifact analysis, and validation conditions compatible with our protocol, but limits dataset breadth. Adding E5 or OpTC would require different label, graph construction, artifact handling, and calibration choices, so we leave them to separate evaluations.

This scope also limits how the semantic signal quality diagnostic should be interpreted. It helps explain the observed E3 pattern, but we do not treat it as a general predictor of benchmark difficulty. Testing whether feature completeness and field entropy predict architectural separation more broadly will require additional datasets with comparable labels and audit depth.

Finally, our comparison evaluates documented configurations under a shared protocol and does not establish the best achievable performance of each architecture. Bounded configuration checks and component sensitivity analyses examine how particular choices affect the results, but further tuning could improve individual systems, including Theseus. Our conclusions are limited to what the audited benchmarks and calibration policy support, leaving model optimality and adversarial robustness outside this scope.

\subsection{Operating Policy}
Our fixed-threshold metrics reflect one conservative deployment policy, where the threshold is set to the maximum benign calibration score. This matches the emphasis on low false positive rates in prior PIDS evaluations~\cite{bilot2025simpler} and keeps calibration independent of the test period. Our threshold sweeps show substantial changes in recovery on some datasets, including Velox on Trace, without closing the Theia gap between Velox and Theseus over the tested range. We interpret fixed-threshold metrics as performance under this conservative operating point, while AP and AUROC measure ranking quality across thresholds and ADP measures attack coverage across thresholds.

\begin{table*}[!t]
    \centering
    \small
    \caption{E3 Benchmark Quality Audit of Raw Collection Streams. Entity linkage tracks event resolution into graph entities. Feature coverage shows the fraction of unique entities with a non-empty attribute. Continuity tracks within-stream timestamp regressions and gaps.}
    \label{tab:benchmark_quality_audit}
    \setlength{\tabcolsep}{5pt}
    \begin{tabular}{l
                    S[table-format=3.1]
                    S[table-format=2.1]
                    S[table-format=2.1]
                    S[table-format=3.1]
                    S[table-format=2.1]
                    S[table-format=2.1]
                    S[table-format=3.1]
                    S[table-format=3.0]
                    S[table-format=1.3]
                    S[table-format=1.3]
                    l}
        \toprule
        & \multicolumn{3}{c}{\textbf{Entity Linkage}} & \multicolumn{4}{c}{\textbf{Feature Coverage (\%)}} & \multicolumn{4}{c}{\textbf{Timestamp Continuity}} \\
        \cmidrule(lr){2-4} \cmidrule(lr){5-8} \cmidrule(lr){9-12}
        \textbf{Dataset} &
        {\makecell[c]{\textbf{Events}\\\textbf{(M)}}} &
        {\makecell[c]{\textbf{Resolved}\\\textbf{(\%)}}} &
        {\makecell[c]{\textbf{Unlinkable}\\\textbf{(\%)}}} &
        {\makecell[c]{\textbf{Proc.}\\\textbf{cmd}}} &
        {\makecell[c]{\textbf{Proc.}\\\textbf{path}}} &
        {\makecell[c]{\textbf{File}\\\textbf{path}}} &
        {\makecell[c]{\textbf{Network}\\\textbf{flow}}} &
        {\makecell[c]{\textbf{Streams}}} &
        {\makecell[c]{\textbf{Gaps $>$60s}\\\textbf{(\%)}}} &
        {\makecell[c]{\textbf{Out-of-order}\\\textbf{(\%)}}} &
        {\textbf{Max gap}} \\
        \midrule
        Cadets   & 41.4  & 88.2 & 11.8 & 99.8  & 0.0  & 13.9 & 100.0 & 10  & 0.000 & 0.001 & 0\,s \\
        FiveDir. & 261.2 & 65.6 & 34.4 & 3.1   & 0.0  & 95.6 & 70.7  & 55  & 0.047 & 6.947 & 8.5 days \\
        Trace    & 813.4 & 41.0 & 59.0 & 100.0 & 0.0  & 98.0 & 100.0 & 204 & 0.000 & 0.071 & 3.1 days \\
        Theia    & 106.0 & 41.9 & 58.1 & 99.6  & 99.6 & 77.7 & 100.0 & 25  & 0.000 & 0.003 & --- \\
        \bottomrule
    \end{tabular}
\end{table*}

\section{Toward Higher-Quality Benchmarks}\label{sec:benchmark_guidance}

Our benchmark analysis identifies five properties needed for reliable architectural comparison: faithful graph construction, complete and diverse semantic attributes, stable temporal structure, documented benign workload variation, and granular labels that match the forensic task. Future benchmarks should treat these as release requirements and document collection conditions, intended uses, and known limitations~\cite{gebru2021datasheets}.

These benchmark properties also motivate testable architectural hypotheses. Future work should evaluate whether combining semantic and structural information improves robustness, ablate semantic attributes and structural features separately to identify the source of performance gains, validate calibration under documented workload shifts, and measure both alerting and process-level recovery.

\subsection{Entity Linkage}
Graph construction loses information when the evaluation schema cannot represent the entity families referenced by raw events. We use the common PIDS abstraction over processes, files, and network flows~\cite{bilot2025simpler,jiang2025orthrus,kairos} so that Theseus and the baselines operate on the same target. Table~\ref{tab:benchmark_quality_audit} reports the fraction of raw events where both subject and object UUIDs resolve under this schema. In Cadets, 11.8\% of raw events are unlinkable, including 10.89\% that lack an object UUID entirely. In Theia, FiveDirections, and Trace, up to 59.0\% of raw events contain an object UUID that does not resolve to a process, file, or network flow. These events may be valid in the raw CDM stream, but their relationships are dropped before detectors receive the data. Benchmark releases and PIDS evaluations should therefore preserve all security-relevant entity families or document and validate the information lost when reducing raw events to an evaluation graph.

\subsection{Semantic Support}
Even when events link correctly, missing attributes limit what a detector can learn from the graph. Some attribute loss is introduced by graph abstraction, but much of it is already present in the raw collection. Table~\ref{tab:benchmark_quality_audit} also reports coverage over unique raw entities before downstream graph construction. Process paths are absent in Cadets, FiveDirections, and Trace. Cadets also has only 130 unique raw process command strings across the full collection. FiveDirections has a broader command vocabulary, but command lines are populated for only 3.1\% of processes. Theia records basic execution metadata much more consistently. Benchmark preparation should prioritize complete command lines, absolute process and file paths, and resolved network endpoints before applying additional graph abstractions. Where the collection supports them, releases should also preserve attributes such as process privileges, file permissions, and event return values, which could provide additional evidence for detection, for example to distinguish failed or unauthorized actions from successful ones.

\subsection{Collection Stability and Temporal Span}
Collection failures can leave gaps and out-of-order timestamps in event streams, distorting temporal splits and graph snapshots before any detector is evaluated. In the released E3 logs (Table~\ref{tab:benchmark_quality_audit}), Cadets has negligible timestamp regressions, whereas FiveDirections has 6.95\% within-stream regressions and gaps of up to 8.5 days, and Trace has gaps of up to 3.1 days. Theia has few regressions, but its documented workload reconfigurations create phase changes in the benign distribution~\cite{liu2025what}. Because current datasets span short periods concentrated around the attacks, these disruptions directly affect the temporal splits. Future releases should include longer captures with documented workload changes and explicit records of collection failures.

\subsection{Support for Robustness Claims}
Robustness to distribution shift requires benchmarks that document multiple benign workload conditions. This allows training and validation to cover different benign behavior without using the final test period. The final evaluation can then include both in-distribution and out-of-distribution test periods, each with benign activity and attacks, to measure whether detection performance generalizes beyond the conditions used for model design and threshold selection.

Adversarial robustness depends on benchmark quality in the same way. On benchmarks with the limitations described above, adaptive attacks may expose dataset weaknesses instead of showing how the detector behaves under realistic adversarial changes. Robustness evaluations should therefore address these limitations before introducing detector-specific adaptive attacks under precise threat models~\cite{arp2022and}.

\subsection{Label Granularity}\label{sec:label_granularity}

\begin{figure}[t]
    \centering
    \definecolor{strictred}{HTML}{C33530}
    \definecolor{sideeffectorange}{HTML}{FF8432}
    \definecolor{benigngreen}{HTML}{6AF190}
    \definecolor{processgreen}{HTML}{C9E8D2}
    \definecolor{nodeyellow}{HTML}{FEF2CC}
    \definecolor{nodeblue}{HTML}{D1D9FC}
    \definecolor{nodegrey}{HTML}{EEEEEE}

    \resizebox{0.95\columnwidth}{!}{
    \begin{tikzpicture}[
        node distance=1.1cm and 1.7cm,
        box/.style={
            rectangle,
            draw=black,
            rounded corners,
            align=center,
            minimum height=0.55cm,
            minimum width=2.05cm,
            font=\ttfamily\small
        },
        proc/.style={box, fill=processgreen},
        file/.style={box, fill=nodeblue},
        sock/.style={box, fill=nodeyellow},
        strict/.style={->, >=Latex, very thick, strictred, font=\scriptsize\sffamily},
        benign/.style={->, >=Latex, thick, benigngreen!60!black, densely dotted, font=\scriptsize\sffamily},
        sideeffect/.style={->, >=Latex, thick, sideeffectorange!85!black, dashed, font=\scriptsize\sffamily}
    ]

    \node[proc] (httpd) {httpd};
    \node[proc, below=of httpd] (sh) {sh};
    \node[proc, below=of sh] (python) {python};
    \node[proc, below=of python] (postgres) {postgres};
    \node[file, below=of postgres] (dump) {dump.sql};
    \node[sock, below=of dump, yshift=0.18cm] (c2) {198.51.100.1};

    \node[file, left=1.35cm of httpd] (lib) {libc.so};
    \node[file, left=1.35cm of sh] (log) {access.log};
    \node[file, left=1.35cm of postgres] (conf) {pg\_hba.conf};

    \node[proc, right=1.95cm of sh] (crash) {payload\_4.9};
    \node[file, right=1.95cm of python, yshift=0.5cm] (shadow) {/etc/shadow};
    \node[sock, right=1.95cm of python, yshift=-0.5cm] (scan) {10.0.0.5};

    \draw[strict] (httpd) -- node[right] {execve} (sh);
    \draw[strict] (sh) -- node[right] {execve} (python);
    \draw[strict] (python) -- node[right] {connect} (postgres);
    \draw[strict] (postgres) -- node[right] {write} (dump);
    \draw[strict] (dump.20) to[out=45,in=-45] node[right] {read} (python.340);
    \draw[strict] (python.200) to[out=240,in=120] node[left] {sendto} (c2.160);

    \draw[benign] (httpd) -- node[above] {read} (lib);
    \draw[benign] (httpd) -- node[above, sloped] {write} (log.north east);
    \draw[benign] (postgres) -- node[above] {read} (conf);

    \draw[sideeffect] (sh) -- node[above] {execve (crash)} (crash);
    \draw[sideeffect] (python) -- node[above, sloped] {open (fail)} (shadow.west);
    \draw[sideeffect] (python) -- node[below, sloped] {connect (fail)} (scan.west);

    \node[
        draw=black!50,
        inner sep=6pt,
        anchor=north west,
        font=\scriptsize\sffamily,
        align=left,
        xshift=-1.2cm,
        yshift=1.1cm
    ] at (scan.south west|-c2.west) {
        \tikz[baseline=-0.5ex]\draw[strict] (0,0) -- (0.6,0); Successful Attack Chain\\[0.5ex]
        \tikz[baseline=-0.5ex]\draw[sideeffect] (0,0) -- (0.6,0); Failed Attempts (Excluded)\\[0.5ex]
        \tikz[baseline=-0.5ex]\draw[benign] (0,0) -- (0.6,0); Benign Execution (Conflated)
    };

    \end{tikzpicture}
    }
    \caption{Illustrative example of evaluation scopes. The current process-level scope typically captures only the successful execution path (solid red). Edge-level annotations can distinguish failed attack-induced actions (dashed orange) from unrelated benign execution (dotted green) that are currently conflated under node-level labeling.}
    \label{fig:causal_scope}
\end{figure}
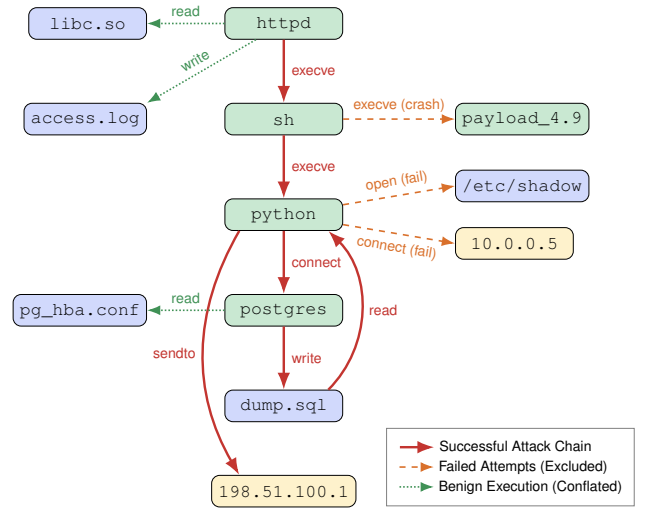

Process-level labels are the only scoring target shared by the four audited E3 datasets and supported by the outputs of all evaluated systems. They are reproducible and match common endpoint detection workflows, but a process label does not indicate which of the process's actions are malicious or when they occur. Fig.~\ref{fig:causal_scope} illustrates the two resulting limitations.

First, a compromised process may continue to perform benign actions unrelated to the attack, while only some of its operations are caused by the intrusion, and a single label conflates both. Second, the scored attack path follows the successful attack, so it can exclude malicious behavior such as failed exploit attempts, persistence, lateral movement, or reconnaissance. A detector should receive credit for surfacing such behavior, while unrelated activity of the same process should not automatically count as malicious.

Our scoring scope builds on the REAPr labels~\cite{reaprgroundtruth} and Liu et al.'s audit~\cite{liu2025what}. Scoring the documented attack path and excluding contaminated nodes avoids rewarding or penalizing detections whose role cannot be established confidently from the DARPA reports, at the cost of the two limitations above.

Finer annotations exist for part of the suite. REAPr provides causal seeds from which one can derive when a process first enters the reconstructed attack subgraph, as well as edge annotations for Cadets, FiveDirections, and Theia, but not for Trace~\cite{reaprgroundtruth}. Using them for comparison would also require a separate evaluation protocol and a mapping from each system's outputs to labeled edges. We therefore retain process labels as the common target across all four datasets.

Where the available evidence supports them, future benchmark releases could complement process labels with edge or event annotations across all datasets, enabling analyses of timing and of individual malicious actions. Such annotations should represent causal ambiguity explicitly and avoid implying perfectly precise attack boundaries.

\section{Conclusion}

PIDS benchmarks can support architectural comparison only when their data and graph abstractions preserve enough relevant signal and their evaluation protocols allow model differences to be measured reliably. Under our auditable methodology and process-level scope, Cadets and Trace still help test whether models can surface attack scenarios, and FiveDirections provides little usable detection signal, but none of the three provides stable evidence for distinguishing architectures or recovering attack-relevant context. On all four audited E3 datasets, an allowlist of executable names and paths achieves the best F1 and MCC at the selected thresholds, so strong operating-point results alone do not establish the value of learned provenance representations.

The experiments also show that alerting and investigation can diverge, as systems can surface attacks while identifying only a small portion of the malicious footprint. Theia, the dataset with the highest semantic signal quality, is the only one on which our reference model clearly improves both ranking and node-level recovery over the other learned systems.

On datasets with limited semantic signal, apparent architectural gains may reflect benchmark-specific cues or calibration choices rather than improved investigation utility. Establishing stronger evidence for model design will require higher-fidelity provenance data and ground truth suited to the evaluation task, alongside metrics that measure alerting and investigation separately. Until then, reported architectural gains must be interpreted within the benchmark conditions and evaluation protocol that produced them.

\section*{Acknowledgments}
We thank the anonymous reviewers and the artifact evaluation committee for their feedback, and Adrien Lafage for helpful discussions on this work. This work was granted access to the HPC resources of IDRIS under allocation 2025-AD011016331 made by GENCI.


\section{Ethical Considerations}

We consider the ethical implications of this study in light of the Menlo Report~\cite{menlo}. The study uses publicly released cybersecurity benchmarks and involves no new user data collection or attempt to identify individuals. Our main ethical concerns are the interpretation of comparative results, fair treatment of prior work, and release of research artifacts.

\textit{Beneficence and risk.}
This study examines conditions under which measured PIDS performance may depend on benchmark artifacts, lexical novelty, or evaluation look-ahead. A main risk is overgeneralization: the results concern the evaluated benchmarks and protocol choices and should not be read as evidence that provenance-based detection is ineffective in general. We distinguish diagnostic evidence from causal conclusions, avoid claims about adversarial robustness beyond the evaluated setting, and do not provide exploit or evasion procedures.

\textit{Justice.}
Re-evaluating published systems can impose reputational costs or additional validation work on their authors and on benchmark maintainers. We reduce this risk by evaluating compatible systems under the same ground-truth, splitting, calibration, and scoring rules and by reporting limitations and uncertainty consistently. We make no claims about researcher intent or competence. Observed differences are attributed to methodological and benchmark conditions, and the released evidence allows others to verify, contest, or extend the findings.

\textit{Respect for law and the public interest.}
We use and redistribute data and derived artifacts under the applicable licenses and release conditions. We did not identify a vulnerability in a deployed system requiring coordinated disclosure. We provide code, processed benchmark files, and checkpoints to support independent verification.

\bibliographystyle{IEEEtran}
\bibliography{ndss2027}

\begin{thebibliography}{10}
\providecommand{\url}[1]{#1}
\csname url@samestyle\endcsname
\providecommand{\newblock}{\relax}
\providecommand{\bibinfo}[2]{#2}
\providecommand{\BIBentrySTDinterwordspacing}{\spaceskip=0pt\relax}
\providecommand{\BIBentryALTinterwordstretchfactor}{4}
\providecommand{\BIBentryALTinterwordspacing}{\spaceskip=\fontdimen2\font plus
\BIBentryALTinterwordstretchfactor\fontdimen3\font minus
  \fontdimen4\font\relax}
\providecommand{\BIBforeignlanguage}[2]{{%
\expandafter\ifx\csname l@#1\endcsname\relax
\typeout{** WARNING: IEEEtran.bst: No hyphenation pattern has been}%
\typeout{** loaded for the language `#1'. Using the pattern for}%
\typeout{** the default language instead.}%
\else
\language=\csname l@#1\endcsname
\fi
#2}}
\providecommand{\BIBdecl}{\relax}
\BIBdecl

\bibitem{zipperle2022provenance}
M.~Zipperle, F.~Gottwalt, E.~Chang, and T.~Dillon, ``Provenance-based intrusion
  detection systems: A survey,'' \emph{ACM Computing Surveys}, vol.~55, no.~7,
  pp. 1--36, 2022.

\bibitem{dong2023we}
F.~Dong, S.~Li, P.~Jiang, D.~Li, H.~Wang, L.~Huang, X.~Xiao, J.~Chen, X.~Luo,
  Y.~Guo \emph{et~al.}, ``Are we there yet? an industrial viewpoint on
  provenance-based endpoint detection and response tools,'' in
  \emph{Proceedings of the 2023 ACM SIGSAC Conference on Computer and
  Communications Security}, 2023, pp. 2396--2410.

\bibitem{sleuth}
M.~N. Hossain, S.~M. Milajerdi, J.~Wang, B.~Eshete, R.~Gjomemo, R.~Sekar,
  S.~Stoller, and V.~Venkatakrishnan, ``{SLEUTH}: Real-time attack scenario
  reconstruction from {COTS} audit data,'' in \emph{26th USENIX Security
  Symposium (USENIX Security 17)}.\hskip 1em plus 0.5em minus 0.4em\relax
  USENIX Association, 2017, pp. 487--504.

\bibitem{holmes}
S.~M. Milajerdi, R.~Gjomemo, B.~Eshete, R.~Sekar, and V.~Venkatakrishnan,
  ``{HOLMES}: Real-time {APT} detection through correlation of suspicious
  information flows,'' in \emph{2019 IEEE Symposium on Security and Privacy
  (SP)}.\hskip 1em plus 0.5em minus 0.4em\relax IEEE, 2019, pp. 1137--1152.

\bibitem{bilot2023gnn}
T.~Bilot, N.~El~Madhoun, K.~Al~Agha, and A.~Zouaoui, ``Graph neural networks
  for intrusion detection: A survey,'' \emph{IEEE Access}, vol.~11, pp.
  49\,114--49\,139, 2023.

\bibitem{kairos}
Z.~Cheng, Q.~Lv, J.~Liang, Y.~Wang, D.~Sun, T.~Pasquier, and X.~Han,
  ``{KAIROS}: Practical intrusion detection and investigation using
  whole-system provenance,'' in \emph{2024 IEEE Symposium on Security and
  Privacy (SP)}.\hskip 1em plus 0.5em minus 0.4em\relax IEEE, 2024.

\bibitem{threatrace}
\BIBentryALTinterwordspacing
S.~Wang, Z.~Wang, T.~Zhou, H.~Sun, X.~Yin, D.~Han, H.~Zhang, X.~Shi, and
  J.~Yang, ``{THREATRACE}: Detecting and tracing host-based threats in node
  level through provenance graph learning,'' \emph{IEEE Transactions on
  Information Forensics and Security}, vol.~17, pp. 3972--3987, 2022. [Online].
  Available: \url{https://doi.org/10.1109/TIFS.2022.3208815}
\BIBentrySTDinterwordspacing

\bibitem{abrar2025reproducibility}
\BIBentryALTinterwordspacing
T.~Abrar, A.~Shamail, M.~J. Iqbal, A.~Ahmed, M.~Abdullah, M.~Shayan, F.~Zaffar,
  T.~Pasquier, D.~Eyers, and A.~Gehani, ``On the reproducibility of
  provenance-based intrusion detection that uses deep learning,'' in
  \emph{Proceedings of the 3rd ACM Conference on Reproducibility and
  Replicability}, ser. ACM REP '25.\hskip 1em plus 0.5em minus 0.4em\relax New
  York, NY, USA: Association for Computing Machinery, 2025, pp. 14--28.
  [Online]. Available: \url{https://doi.org/10.1145/3736731.3746140}
\BIBentrySTDinterwordspacing

\bibitem{bilot2025simpler}
T.~Bilot, B.~Jiang, Z.~Li, N.~El~Madhoun, K.~Al~Agha, A.~Zouaoui, and
  T.~Pasquier, ``Sometimes simpler is better: A comprehensive analysis of
  state-of-the-art provenance-based intrusion detection systems,'' in
  \emph{34th USENIX Security Symposium (USENIX Security 25)}, 2025, pp.
  7193--7212.

\bibitem{liu2025what}
\BIBentryALTinterwordspacing
J.~Liu, M.~A. Inam, A.~Goyal, A.~Riddle, K.~Westfall, and A.~Bates, ``What we
  talk about when we talk about logs: Understanding the effects of dataset
  quality on endpoint threat detection research,'' in \emph{2025 IEEE Symposium
  on Security and Privacy (SP)}, 2025, pp. 112--129. [Online]. Available:
  \url{https://doi.org/10.1109/SP61157.2025.00112}
\BIBentrySTDinterwordspacing

\bibitem{michael2020forensic}
\BIBentryALTinterwordspacing
N.~Michael, J.~Mink, J.~Liu, S.~Gaur, W.~U. Hassan, and A.~Bates, ``On the
  forensic validity of approximated audit logs,'' in \emph{Proceedings of the
  36th Annual Computer Security Applications Conference}, ser. ACSAC '20.\hskip
  1em plus 0.5em minus 0.4em\relax New York, NY, USA: Association for Computing
  Machinery, 2020, pp. 189--202. [Online]. Available:
  \url{https://doi.org/10.1145/3427228.3427272}
\BIBentrySTDinterwordspacing

\bibitem{geirhos2020shortcut}
R.~Geirhos, J.-H. Jacobsen, C.~Michaelis, R.~Zemel, W.~Brendel, M.~Bethge, and
  F.~A. Wichmann, ``Shortcut learning in deep neural networks,'' \emph{Nature
  Machine Intelligence}, vol.~2, no.~11, pp. 665--673, 2020.

\bibitem{jiang2025orthrus}
B.~Jiang, T.~Bilot, N.~El~Madhoun, K.~Al~Agha, A.~Zouaoui, S.~Iqbal, X.~Han,
  and T.~Pasquier, ``{ORTHRUS}: Achieving high quality of attribution in
  provenance-based intrusion detection systems,'' in \emph{34th USENIX Security
  Symposium (USENIX Security 25)}, 2025, pp. 7173--7192.

\bibitem{hassan2019nodoze}
W.~U. Hassan, S.~Guo, D.~Li, Z.~Chen, K.~Jee, Z.~Li, and A.~Bates, ``{NoDoze}:
  Combatting threat alert fatigue with automated provenance triage,'' in
  \emph{26th Annual Network and Distributed System Security Symposium (NDSS)},
  2019.

\bibitem{fang2022back}
P.~Fang, P.~Gao, C.~Liu, E.~Ayday, K.~Jee, T.~Wang, Y.~F. Ye, Z.~Liu, and
  X.~Xiao, ``{Back-Propagating} system dependency impact for attack
  investigation,'' in \emph{31st USENIX Security Symposium (USENIX Security
  22)}.\hskip 1em plus 0.5em minus 0.4em\relax Boston, MA: USENIX Association,
  Aug. 2022, pp. 2461--2478.

\bibitem{li2024nodlink}
S.~Li, F.~Dong, X.~Xiao, H.~Wang, F.~Shao, J.~Chen, Y.~Guo, X.~Chen, and D.~Li,
  ``{NODLINK}: An online system for fine-grained {APT} attack detection and
  investigation,'' in \emph{31st Annual Network and Distributed System Security
  Symposium (NDSS)}, 2024.

\bibitem{arp2022and}
D.~Arp, E.~Quiring, F.~Pendlebury, A.~Warnecke, F.~Pierazzi, C.~Wressnegger,
  L.~Cavallaro, and K.~Rieck, ``Dos and don'ts of machine learning in computer
  security,'' in \emph{31st USENIX Security Symposium (USENIX Security 22)},
  2022, pp. 3971--3988.

\bibitem{darpa-tc-e3}
{DARPA}, ``Transparent computing engagement 3 data release,''
  \url{https://github.com/darpa-i2o/Transparent-Computing/blob/master/README-E3.md},
  2018, commit: \texttt{e94c9f2}. Accessed: December 2025.

\bibitem{reaprgroundtruth}
J.~Liu, M.~A. Inam, A.~Goyal, K.~Westfall, A.~Riddle, and A.~Bates, ``{REAPr}:
  Recovery every attack process,''
  \url{https://bitbucket.org/sts-lab/reapr-ground-truth}, 2023, commit:
  \texttt{e726c01}. Accessed: July 2026.

\bibitem{riddle2023atlasv2}
\BIBentryALTinterwordspacing
A.~Riddle, K.~Westfall, and A.~Bates, ``{ATLASv2}: {ATLAS} attack engagements,
  version 2,'' 2023. [Online]. Available:
  \url{https://arxiv.org/abs/2401.01341}
\BIBentrySTDinterwordspacing

\bibitem{liu2026windows}
J.~Liu, M.~A. Inam, A.~Goyal, D.~Greenenwald, A.~Bates, and S.~Chittal, ``How
  to effectively trace provenance on {Windows} endpoint detection \& response
  telemetry,'' in \emph{Workshop on Attack Provenance, Reasoning, and
  Investigation for Security in the Monitored Environment (PRISM'26)}, 2026.

\bibitem{spade}
A.~Gehani and D.~Tariq, ``{SPADE}: Support for provenance auditing in
  distributed environments,'' in \emph{ACM/IFIP/USENIX International Conference
  on Distributed Systems Platforms and Open Distributed Processing}.\hskip 1em
  plus 0.5em minus 0.4em\relax Springer, 2012, pp. 101--120.

\bibitem{camflow}
T.~Pasquier, X.~Han, M.~Goldstein, T.~Moyer, D.~Eyers, M.~Seltzer, and
  J.~Bacon, ``Practical whole-system provenance capture,'' in \emph{Proceedings
  of the ACM Symposium on Cloud Computing (SoCC '17)}.\hskip 1em plus 0.5em
  minus 0.4em\relax ACM, 2017.

\bibitem{bates2015whole-system}
A.~Bates, D.~J. Tian, K.~R. Butler, and T.~Moyer, ``Trustworthy {Whole-System}
  provenance for the {Linux} kernel,'' in \emph{24th USENIX Security Symposium
  (USENIX Security 15)}.\hskip 1em plus 0.5em minus 0.4em\relax Washington,
  D.C.: USENIX Association, Aug. 2015, pp. 319--334.

\bibitem{lee2013loggc}
K.~H. Lee, X.~Zhang, and D.~Xu, ``{LogGC}: Garbage collecting audit log,'' in
  \emph{Proceedings of the 2013 ACM SIGSAC Conference on Computer and
  Communications Security}, 2013, pp. 1005--1016.

\bibitem{beep}
------, ``High accuracy attack provenance via binary-based execution
  partition,'' in \emph{20th Annual Network and Distributed System Security
  Symposium (NDSS)}, vol.~16, 2013.

\bibitem{xu2016reduction}
\BIBentryALTinterwordspacing
Z.~Xu, Z.~Wu, Z.~Li, K.~Jee, J.~Rhee, X.~Xiao, F.~Xu, H.~Wang, and G.~Jiang,
  ``High fidelity data reduction for big data security dependency analyses,''
  in \emph{Proceedings of the 2016 ACM SIGSAC Conference on Computer and
  Communications Security}, ser. CCS '16.\hskip 1em plus 0.5em minus
  0.4em\relax New York, NY, USA: Association for Computing Machinery, 2016, pp.
  504--516. [Online]. Available: \url{https://doi.org/10.1145/2976749.2978378}
\BIBentrySTDinterwordspacing

\bibitem{liu2018towards}
Y.~Liu, M.~Zhang, D.~Li, K.~Jee, Z.~Li, Z.~Wu, J.~Rhee, and P.~Mittal,
  ``Towards a timely causality analysis for enterprise security,'' in
  \emph{25th Annual Network and Distributed System Security Symposium (NDSS)},
  2018.

\bibitem{zeng2021watson}
J.~Zeng, Z.~L. Chua, Y.~Chen, K.~Ji, Z.~Liang, and J.~Mao, ``{WATSON}:
  Abstracting behaviors from audit logs via aggregation of contextual
  semantics,'' in \emph{28th Annual Network and Distributed System Security
  Symposium (NDSS)}, 2021.

\bibitem{deeplog}
M.~Du, F.~Li, G.~Zheng, and V.~Srikumar, ``{DeepLog}: Anomaly detection and
  diagnosis from system logs through deep learning,'' in \emph{Proceedings of
  the 2017 ACM SIGSAC Conference on Computer and Communications
  Security}.\hskip 1em plus 0.5em minus 0.4em\relax ACM, 2017, pp. 1285--1298.

\bibitem{guo2021logbert}
H.~Guo, S.~Yuan, and X.~Wu, ``{LogBERT}: Log anomaly detection via {BERT},'' in
  \emph{2021 International Joint Conference on Neural Networks (IJCNN)}.\hskip
  1em plus 0.5em minus 0.4em\relax IEEE, 2021, pp. 1--8.

\bibitem{unicorn}
\BIBentryALTinterwordspacing
X.~Han, T.~Pasquier, A.~Bates, J.~Mickens, and M.~Seltzer, ``{Unicorn}: Runtime
  provenance-based detector for advanced persistent threats,'' in \emph{27th
  Network and Distributed System Security Symposium (NDSS)}, 2020. [Online].
  Available: \url{https://doi.org/10.14722/ndss.2020.24046}
\BIBentrySTDinterwordspacing

\bibitem{wang2020provdetector}
Q.~Wang, W.~U. Hassan, D.~Li, K.~Jee, X.~Yu, K.~Zou, J.~Rhee, Z.~Chen,
  W.~Cheng, C.~A. Gunter, and H.~Chen, ``You are what you do: Hunting stealthy
  malware via data provenance analysis,'' in \emph{27th Annual Network and
  Distributed System Security Symposium (NDSS)}, 2020.

\bibitem{jia2024magic}
Z.~Jia, Y.~Xiong, Y.~Nan, Y.~Zhang, J.~Zhao, and M.~Wen, ``{MAGIC}: Detecting
  advanced persistent threats via masked graph representation learning,'' in
  \emph{33rd USENIX Security Symposium, USENIX Security 2024}, 2024.

\bibitem{flash}
M.~U. Rehman, H.~Ahmadi, and W.~U. Hassan, ``{Flash}: A comprehensive approach
  to intrusion detection via provenance graph representation learning,'' in
  \emph{2024 IEEE Symposium on Security and Privacy (SP)}.\hskip 1em plus 0.5em
  minus 0.4em\relax IEEE, 2024, pp. 3552--3570.

\bibitem{rcaid}
\BIBentryALTinterwordspacing
A.~Goyal, G.~Wang, and A.~Bates, ``{R-CAID}: Embedding root cause analysis
  within provenance-based intrusion detection,'' in \emph{2024 IEEE Symposium
  on Security and Privacy (SP)}, 2024, pp. 3515--3532. [Online]. Available:
  \url{https://doi.org/10.1109/SP54263.2024.00253}
\BIBentrySTDinterwordspacing

\bibitem{davis2006relationship}
J.~Davis and M.~Goadrich, ``The relationship between {Precision-Recall} and
  {ROC} curves,'' in \emph{Proceedings of the 23rd International Conference on
  Machine Learning}, 2006, pp. 233--240.

\bibitem{saito2015precision}
T.~Saito and M.~Rehmsmeier, ``The precision-recall plot is more informative
  than the {ROC} plot when evaluating binary classifiers on imbalanced
  datasets,'' \emph{PloS one}, vol.~10, no.~3, p. e0118432, 2015.

\bibitem{chicco2020advantages}
D.~Chicco and G.~Jurman, ``The advantages of the {Matthews} correlation
  coefficient ({MCC}) over {F1} score and accuracy in binary classification
  evaluation,'' \emph{BMC genomics}, vol.~21, no.~1, p.~6, 2020.

\bibitem{kan2025tesseract}
\BIBentryALTinterwordspacing
Z.~Kan, S.~McFadden, D.~Arp, F.~Pendlebury, R.~Jordaney, J.~Kinder,
  F.~Pierazzi, and L.~Cavallaro, ``{TESSERACT}: Eliminating experimental bias
  in malware classification across space and time (extended version),'' 2025.
  [Online]. Available: \url{https://arxiv.org/abs/2402.01359}
\BIBentrySTDinterwordspacing

\bibitem{pragmatic-assessment}
G.~Apruzzese, P.~Laskov, and J.~Schneider, ``{SoK}: Pragmatic assessment of
  machine learning for network intrusion detection,'' in \emph{2023 IEEE 8th
  European Symposium on Security and Privacy (EuroS\&P)}.\hskip 1em plus 0.5em
  minus 0.4em\relax IEEE, 2023, pp. 592--614.

\bibitem{reimers2017reporting}
N.~Reimers and I.~Gurevych, ``Reporting score distributions makes a difference:
  Performance study of {LSTM}-networks for sequence tagging,'' in
  \emph{Proceedings of the 2017 Conference on Empirical Methods in Natural
  Language Processing}, 2017, pp. 338--348.

\bibitem{bouthillier2021accounting}
\BIBentryALTinterwordspacing
X.~Bouthillier, P.~Delaunay, M.~Bronzi, A.~Trofimov, B.~Nichyporuk, J.~Szeto,
  N.~Mohammadi~Sepahvand, E.~Raff, K.~Madan, V.~Voleti, S.~Ebrahimi~Kahou,
  V.~Michalski, T.~Arbel, C.~Pal, G.~Varoquaux, and P.~Vincent, ``Accounting
  for variance in machine learning benchmarks,'' in \emph{Proceedings of
  Machine Learning and Systems}, vol.~3, 2021, pp. 747--769. [Online].
  Available:
  \url{https://proceedings.mlsys.org/paper_files/paper/2021/file/0184b0cd3cfb185989f858a1d9f5c1eb-Paper.pdf}
\BIBentrySTDinterwordspacing

\bibitem{atlas}
A.~Alsaheel, Y.~Nan, S.~Ma, L.~Yu, G.~Walkup, Z.~B. Celik, X.~Zhang, and D.~Xu,
  ``{ATLAS}: A sequence-based learning approach for attack investigation,'' in
  \emph{30th USENIX Security Symposium (USENIX Security 21)}.\hskip 1em plus
  0.5em minus 0.4em\relax USENIX Association, Aug. 2021, pp. 3005--3022.

\bibitem{majorczyk2025newhope}
\BIBentryALTinterwordspacing
F.~Majorczyk, B.~Pilastre, and F.~Dijoud, ``A new hope for {DARPA} {OpTC},'' in
  \emph{2025 Annual Computer Security Applications Conference Workshops (ACSAC
  Workshops)}.\hskip 1em plus 0.5em minus 0.4em\relax IEEE, 2025, pp. 551--561.
  [Online]. Available: \url{https://doi.org/10.1109/ACSACW69556.2025.00064}
\BIBentrySTDinterwordspacing

\bibitem{shannon1948mathematical}
C.~E. Shannon, ``A mathematical theory of communication,'' \emph{The Bell
  system technical journal}, vol.~27, no.~3, pp. 379--423, 1948.

\bibitem{barr2021survivalism}
F.~Barr-Smith, X.~Ugarte-Pedrero, M.~Graziano, R.~Spolaor, and I.~Martinovic,
  ``Survivalism: Systematic analysis of {Windows} malware
  living-off-the-land,'' in \emph{2021 IEEE Symposium on Security and Privacy
  (SP)}.\hskip 1em plus 0.5em minus 0.4em\relax IEEE, 2021, pp. 1557--1574.

\bibitem{guerra2025selfsupervised}
\BIBentryALTinterwordspacing
L.~Guerra, T.~Chapuis, G.~Duc, P.~Mozharovskyi, and V.-T. Nguyen,
  ``Self-supervised learning of graph representations for network intrusion
  detection,'' in \emph{Advances in Neural Information Processing Systems},
  vol.~38.\hskip 1em plus 0.5em minus 0.4em\relax Curran Associates, Inc.,
  2025, pp. 109\,471--109\,501. [Online]. Available:
  \url{https://doi.org/10.52202/085713-3653}
\BIBentrySTDinterwordspacing

\bibitem{lo2022egraphsage}
\BIBentryALTinterwordspacing
W.~W. Lo, S.~Layeghy, M.~Sarhan, M.~Gallagher, and M.~Portmann,
  ``{E-GraphSAGE}: A graph neural network based intrusion detection system for
  {IoT},'' in \emph{NOMS 2022-2022 IEEE/IFIP Network Operations and Management
  Symposium}, 2022, pp. 1--9. [Online]. Available:
  \url{https://doi.org/10.1109/NOMS54207.2022.9789878}
\BIBentrySTDinterwordspacing

\bibitem{vaswani2017attention}
A.~Vaswani, N.~Shazeer, N.~Parmar, J.~Uszkoreit, L.~Jones, A.~N. Gomez,
  {\L}.~Kaiser, and I.~Polosukhin, ``Attention is all you need,'' in
  \emph{Advances in Neural Information Processing Systems (NeurIPS)}, vol.~30,
  2017, pp. 5998--6008.

\bibitem{mikolov2013efficient}
T.~Mikolov, K.~Chen, G.~Corrado, and J.~Dean, ``Efficient estimation of word
  representations in vector space,'' \emph{arXiv preprint arXiv:1301.3781},
  2013.

\bibitem{pidsmaker-repo}
T.~Bilot \emph{et~al.}, ``{PIDSMaker}: An {ML} framework for building
  provenance-based intrusion detection systems,''
  \url{https://github.com/ubc-provenance/PIDSMaker}, 2025, commit:
  \texttt{2a73886}. Accessed: December 2025.

\bibitem{magic-repo}
Z.~Jia \emph{et~al.}, ``{MAGIC}: Official implementation,''
  \url{https://github.com/FDUDSDE/MAGIC}, 2024, commit: \texttt{aa0b647}.
  Accessed: December 2025.

\bibitem{pytorch}
A.~Paszke, S.~Gross, F.~Massa, A.~Lerer, J.~Bradbury, G.~Chanan, T.~Killeen,
  Z.~Lin, N.~Gimelshein, L.~Antiga \emph{et~al.}, ``{PyTorch}: An imperative
  style, high-performance deep learning library,'' in \emph{Advances in Neural
  Information Processing Systems (NeurIPS)}, vol.~32, 2019.

\bibitem{pyg}
\BIBentryALTinterwordspacing
M.~Fey and J.~E. Lenssen, ``Fast graph representation learning with {PyTorch
  Geometric},'' in \emph{ICLR Workshop on Representation Learning on Graphs and
  Manifolds}, 2019. [Online]. Available: \url{https://arxiv.org/abs/1903.02428}
\BIBentrySTDinterwordspacing

\bibitem{loshchilov2018decoupled}
\BIBentryALTinterwordspacing
I.~Loshchilov and F.~Hutter, ``Decoupled weight decay regularization,'' in
  \emph{International Conference on Learning Representations}, 2019. [Online].
  Available: \url{https://openreview.net/forum?id=Bkg6RiCqY7}
\BIBentrySTDinterwordspacing

\bibitem{gebru2021datasheets}
T.~Gebru, J.~Morgenstern, B.~Vecchione, J.~W. Vaughan, H.~Wallach,
  H.~Daum{\'e}~III, and K.~Crawford, ``Datasheets for datasets,''
  \emph{Communications of the ACM}, vol.~64, no.~12, pp. 86--92, 2021.

\bibitem{menlo}
D.~Dittrich and E.~Kenneally, ``The {Menlo} {Report}: Ethical principles
  guiding information and communication technology research,'' U.S. Department
  of Homeland Security, Tech. Rep., 2012.

\bibitem{manzoor2016streamspot}
\BIBentryALTinterwordspacing
E.~Manzoor, S.~M. Milajerdi, and L.~Akoglu, ``Fast memory-efficient anomaly
  detection in streaming heterogeneous graphs,'' in \emph{Proceedings of the
  22nd ACM SIGKDD International Conference on Knowledge Discovery and Data
  Mining}, ser. KDD '16.\hskip 1em plus 0.5em minus 0.4em\relax New York, NY,
  USA: Association for Computing Machinery, 2016, pp. 1035--1044. [Online].
  Available: \url{https://doi.org/10.1145/2939672.2939783}
\BIBentrySTDinterwordspacing

\end{thebibliography}

\appendices

\section{Open Science}\label{app:openscience}

To support reproducibility and future work, we publicly release the implementation and adapted baseline evaluation code at \url{https://github.com/lorenzo9uerra/theseus}. The processed DARPA TC E3 node and event tables are archived at \url{https://doi.org/10.5281/zenodo.18450779}, while the evaluation artifacts are archived at \url{https://doi.org/10.5281/zenodo.22343430}. The latter include graph caches, model checkpoints, retained baseline artifacts, sanitized evaluation logs, and the processed ATLASv2 files and labels used in the secondary evaluation. The repository also provides a \texttt{uv.lock} file and a container recipe to improve environment reproducibility across machines and clusters.

The released code identifies the upstream versions on which our ground truth and baseline adaptations are based: REAPr commit \texttt{e726c01}, Magic commit \texttt{aa0b647}, and PIDSMaker commit \texttt{2a73886}. Our repository contains the modifications required to integrate the baselines with the shared data and evaluation pipeline, while preserving their model logic. The released artifacts contain the exact ground-truth files used in our evaluation.

\section{Protocol Details and Secondary Diagnostics}

\subsection{Benchmark Viability}\label{app:benchmark_viability}

\begin{table*}[t]
\centering
\footnotesize
\setlength{\tabcolsep}{3pt}
\caption{Benchmark eligibility under our evaluation protocol. Primary datasets require compatible process labels, public audit support for artifacts and ambiguous nodes, and independent attack data for validation.}
\label{tab:benchmark_selection}
\begin{tabular}{p{2.5cm} c c c l p{8cm}}
\toprule
\makecell[l]{\textbf{Dataset}} &
\makecell[c]{\textbf{Compatible}\\\textbf{labels}} &
\makecell[c]{\textbf{Dataset audit}\\\textbf{support}} &
\makecell[c]{\textbf{Attack}\\\textbf{validation}} &
\makecell[l]{\textbf{Role}} &
\makecell[l]{\textbf{Rationale}} \\
\midrule
E3-Cadets &
\yesmark & \yesmark & \yesmark & Primary &
Meets the label, audit, and validation requirements. \\

E3-FiveDirections &
\yesmark & \yesmark & \yesmark & Primary &
Meets the label, audit, and validation requirements. \\

E3-Trace &
\yesmark & \yesmark & \yesmark & Primary &
Meets the label, audit, and validation requirements. \\

E3-Theia &
\yesmark & \yesmark & \yesmark & Primary &
Meets the label, audit, and validation requirements. \\

E3-ClearScope &
\nomark & \nomark & \nomark & Excluded &
REAPr provides no compatible process labels for ClearScope, whose timestamps use arbitrary units~\cite{reaprgroundtruth}. \\

ATLASv2 &
\yesmark & \yesmark & \nomark & Secondary &
REAPr process UUIDs map directly to reconstructed nodes, but each host has only one attack window, so we calibrate thresholds on benign training scores. \\

E5 (Cadets / Theia / ClearScope) &
\nomark & \nomark & \variesmark & Excluded &
We found no public audit documenting artifact handling, treatment of ambiguous nodes, and scoring exclusions for the labels used in prior evaluations. Unlike Cadets and ClearScope, the Theia E5 subset used in those evaluations contains only one labeled attack scenario~\cite{bilot2025simpler}. \\

OpTC (H051 / H201 / H501) &
\nomark & \yesmark & \nomark & Excluded &
Available labels require reconciliation with our process scoring target, and each selected host has only one attack scenario. \\

StreamSpot~\cite{manzoor2016streamspot} / Unicorn Wget~\cite{unicorn} &
\nomark & \nomark & \nomark & Excluded &
Graph-level labels do not support the common process scoring target. \\
\bottomrule
\end{tabular}
\end{table*}

Table~\ref{tab:benchmark_selection} summarizes dataset eligibility under the primary protocol, with ATLASv2 providing secondary evidence alongside E3.

\paragraph{E5}
E5 has been used in prior broad PIDS evaluations~\cite{jiang2025orthrus,bilot2025simpler}, but we are not aware of public audit support documenting artifact handling, ambiguous node treatment, and scoring exclusions for the labels used in those evaluations. Its support for attack validation also varies by host: unlike Cadets and ClearScope, the Theia E5 subset used in those evaluations contains only one labeled attack scenario. Including E5 would therefore require a separate audit of the labels and artifact treatment before it could be evaluated under the same controlled protocol as audited E3.

\paragraph{OpTC}
OpTC is better documented after the correction and labeling work of Majorczyk et al.~\cite{majorczyk2025newhope}, but its available labels define a different scoring target from the REAPr process-level labels used in our E3 experiments. Majorczyk et al.'s pipeline reconstructs malicious host events and network flows from OpTC-specific evidence, including Red Team descriptions, agent PIDs, C\&C infrastructure, process-child propagation, and host/network correlation. These labels are not directly equivalent to the REAPr process labels combined with Liu et al.'s artifact analysis and scoring exclusions~\cite{reaprgroundtruth,liu2025what}.

A corrected OpTC comparison would require a separate graph construction and label reconciliation procedure, followed by an audit of how those choices affect each baseline. Each selected OpTC host benchmark exposes only one attack scenario, so it does not support the independent validation split containing attack data used in our primary regime without either omitting attack data from validation or splitting the same attack campaign. We leave corrected OpTC to a separate evaluation in future work.

\subsection{Dataset Splits}

\begin{table}[t]
\centering
\small
\caption{Temporal data splits used for evaluation. E3 dates refer to April 2018 and ATLASv2 dates to July 2022. Every development day precedes the test period. Attack days are shown to clarify the split design.}\label{tab:splits}
\begin{tabular}{lcccc}
\toprule
\textbf{Dataset} &
\makecell{\textbf{Attack}\\\textbf{Days}} &
\makecell{\textbf{Training}\\\textbf{Days}} &
\makecell{\textbf{Validation}\\\textbf{Days}} &
\makecell{\textbf{Test}\\\textbf{Days}} \\
\midrule
Cadets & 6, 11--13 & 2--5, 7--9 & 6, 10 & 11--13 \\
FiveDir. & 9, 11--13 & 2--7 & 8, 9 & 10--13 \\
Trace & 10, 12, 13 & 9, 11 & 10 & 12, 13 \\
Theia (T3) & 10, 13 & 11 & 10 & 12, 13 \\
ATLASv2-h1 & 19--20 & 15--18 & --- & 19--20 \\
ATLASv2-h2 & 19--20 & 15--18 & --- & 19--20 \\
\bottomrule
\end{tabular}
\end{table}

Table~\ref{tab:splits} summarizes the fixed E3 and ATLASv2 partitions. E3 calendar days use US Eastern time consistently for day assignment and interval boundaries. All development data precede the final test period, although training and validation days may be interleaved within the pre-test period when validation data containing attacks are limited. Training and validation days remain disjoint.

Following Liu et al.~\cite{liu2025what}, we restrict Theia to the stable T3 period. Within T3, Day~10 is the only pre-test day with attack activity, so it serves as the validation day, with Day~11 used for training and Days~12--13 for testing. Trace similarly uses Day~10 for validation, Days~9 and~11 for training, and Days~12--13 for testing. ATLASv2 has no validation split because July~19--20 form a single engagement window containing attacks with similar payload structure.

\subsection{Baseline Configuration Checks}\label{app:baseline_checks}
For the learned baselines, we ran bounded sensitivity sweeps around the released or adapted configurations for each primary dataset. For Magic, we varied learning rate, weight decay, mask rate, and $\alpha_l$. For Velox, we varied learning rate, weight decay, and output dimension. For Orthrus, we varied learning rate, weight decay, dropout, hidden and output dimensions, and the temporal neighborhood size where feasible. On Trace and FiveDirections, the dense graph topology made larger temporal neighborhoods infeasible within the shared resource budget, with runs exceeding 128\,GB of RAM. The sweeps produced local variation in individual metrics, but none changed the qualitative interpretation of the experiments. We report the released or adapted configurations as the primary baseline settings, using these bounded sweeps to check sensitivity to nearby settings.

\subsection{Command-Line Novelty as an Allowlist Key}\label{app:command_allowlist}

\begin{figure*}[t]
    \centering
    \subfloat[Theseus]{%
        \includegraphics[width=0.49\textwidth]{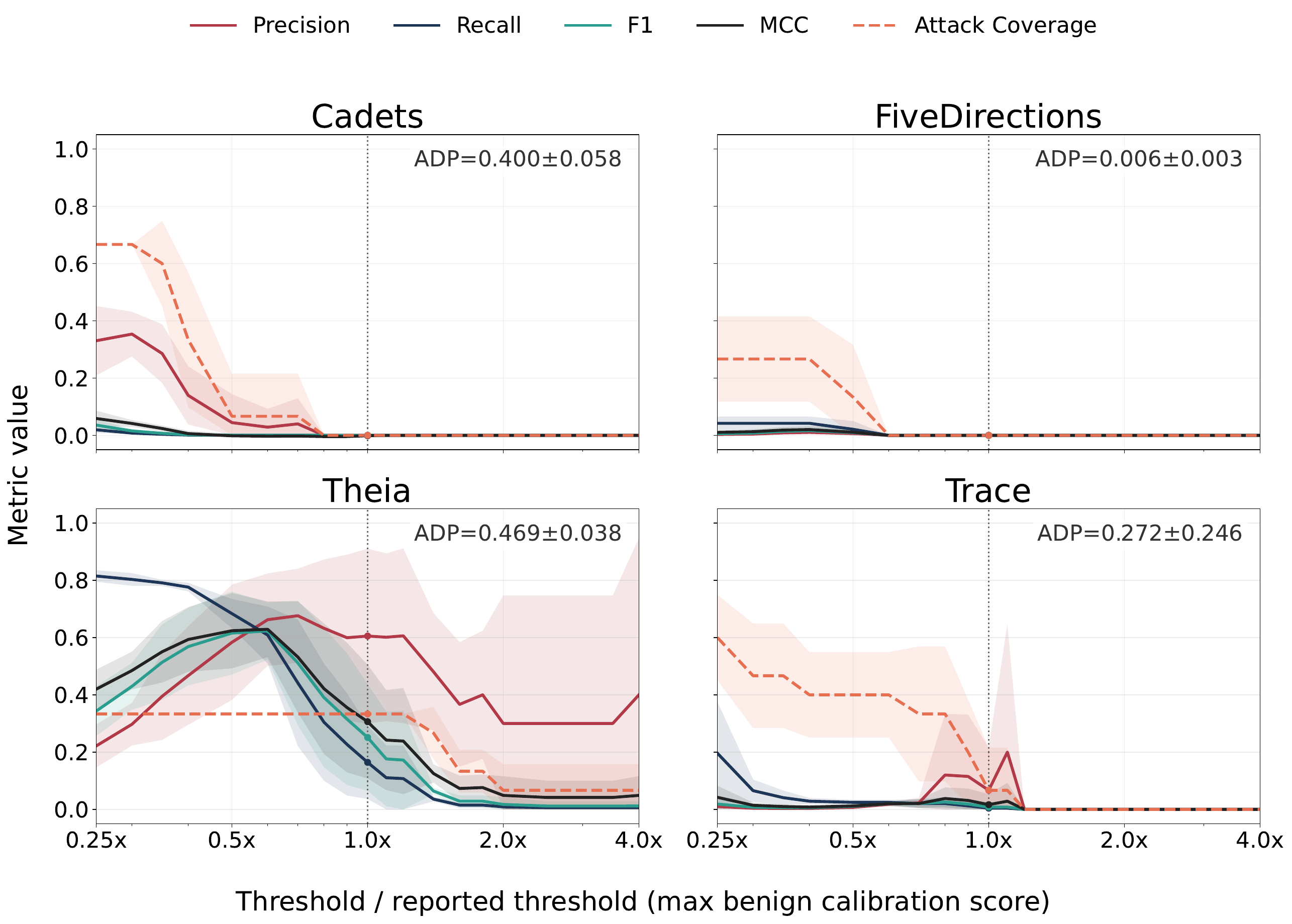}%
    }
    \hfill
    \subfloat[Velox]{%
        \includegraphics[width=0.49\textwidth]{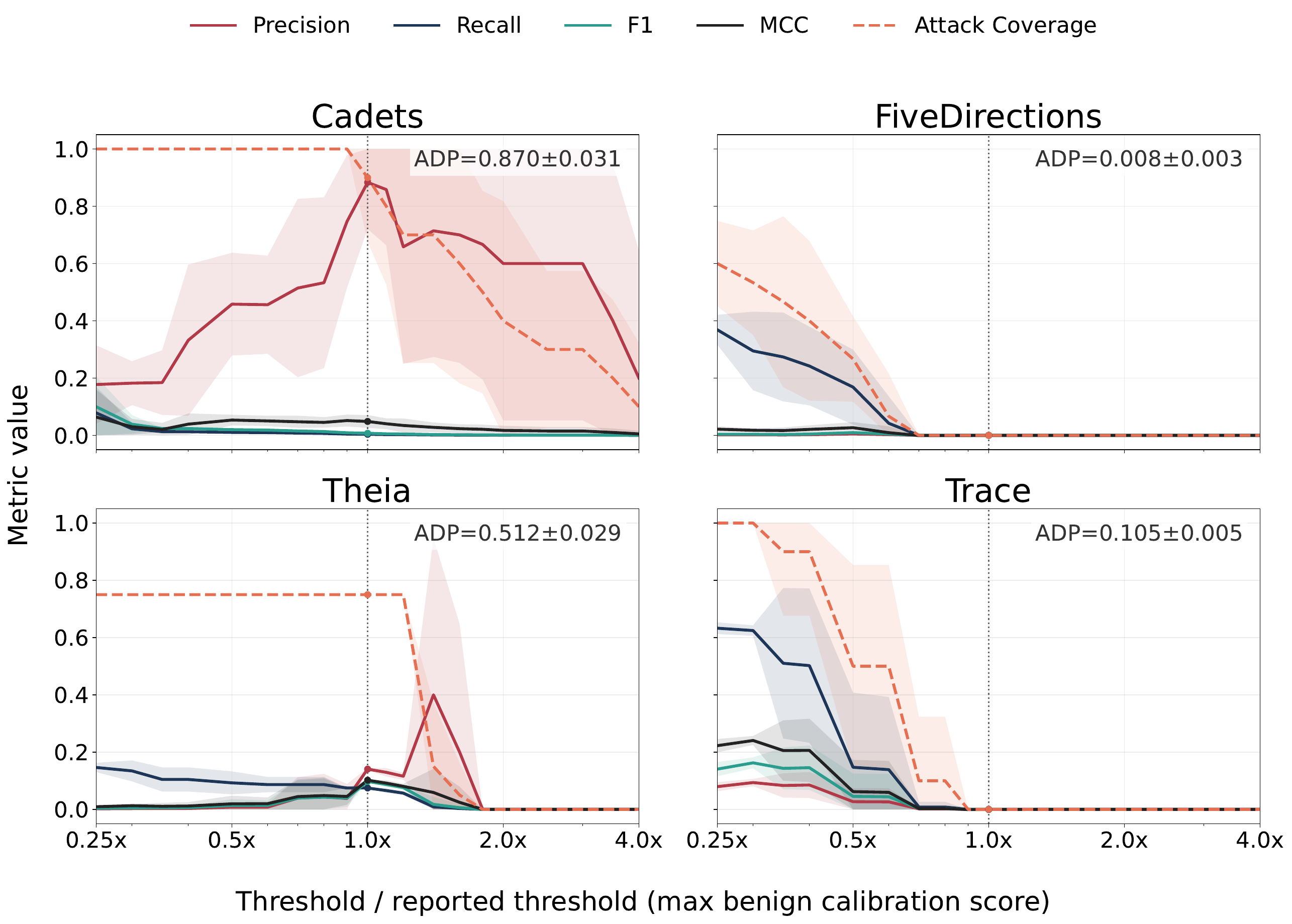}%
    }
    \caption{Threshold sensitivity on DARPA TC E3 for frozen Theseus and Velox models. Curves and annotated ADP values are means over five seeds, and shaded bands show one standard deviation. The horizontal axis shows the threshold relative to the reported threshold, set to the maximum benign validation score, with the dotted line marking \(1.0\times\).}
    \label{fig:threshold_sensitivity}
\end{figure*}

To test whether richer textual fields strengthen the allowlist diagnostic, we replace executable/path novelty with exact command-line novelty, keeping the same benign reference data and process labels. As Table~\ref{tab:command_allowlist} shows, the richer key flags many more processes, generally raising recall but introducing many false positives. Cadets is unchanged, since its command-line field adds no further distinction. On FiveDirections, command-line novelty flags more than seven times as many processes and lowers precision and F1. On Theia and Trace, FPR climbs to 0.0919 and 0.2765, sharply reducing precision, F1, and MCC. Overall, the exact command line is not a stronger allowlist key than executable names and paths.

\begin{table}[t]
\centering
\small
\setlength{\tabcolsep}{5pt}
\caption{Executable/path and command-line novelty on audited E3. Bold values mark the better result within each dataset, with lower values preferred for FPR.}
\label{tab:command_allowlist}
\begin{tabular}{llrrrr}
\toprule
\textbf{Dataset} &
\textbf{Novelty key} &
\textbf{Flagged} &
\textbf{Prec.} &
\textbf{Recall} &
\textbf{FPR} \\
\midrule
\multirow{2}{*}{Cadets}
& Executable/path & 36 & \textbf{0.4286} & \textbf{0.0105} & \textbf{0.0012} \\
& Command line & 36 & \textbf{0.4286} & \textbf{0.0105} & \textbf{0.0012} \\
\midrule
\multirow{2}{*}{FiveDir.}
& Executable/path & 322 & \textbf{0.0031} & 0.0385 & \textbf{0.0014} \\
& Command line & 2{,}287 & 0.0013 & \textbf{0.1154} & 0.0102 \\
\midrule
\multirow{2}{*}{Theia}
& Executable/path & 226 & \textbf{0.1770} & 0.5970 & \textbf{0.0024} \\
& Command line & 7{,}167 & 0.0059 & \textbf{0.6269} & 0.0919 \\
\midrule
\multirow{2}{*}{Trace}
& Executable/path & 700 & \textbf{0.0573} & 0.8163 & \textbf{0.0003} \\
& Command line & 690{,}666 & 0.0001 & \textbf{0.9796} & 0.2765 \\
\bottomrule
\end{tabular}
\end{table}

\subsection{Threshold Sensitivity}\label{app:threshold_sensitivity}
Using frozen checkpoints, we vary only the threshold multiplier relative to the maximum benign validation score. Fig.~\ref{fig:threshold_sensitivity} shows the full E3 curves for Theseus and Velox.

\section{Theseus Details}

\subsection{Attention Masking Sensitivity}\label{sec:mask_ratio}

\begin{table}[t]
\centering
\footnotesize
\setlength{\tabcolsep}{5pt}
\caption{Attention masking sensitivity on Theia under the main conservative threshold policy. Values are means and standard deviations over five seeds. Bold values mark the best displayed mean.}\label{tab:mask_ablation}
\begin{tabular}{lccc}
\toprule
\textbf{Mask Ratio} & \textbf{AP $\uparrow$} & \textbf{MCC $\uparrow$} & \textbf{FPR $\downarrow$} \\
\midrule
0.00 & \bres{0.605}{0.118} & \res{0.423}{0.196} & \bres{0.00004}{0.00003} \\
0.10 & \res{0.598}{0.079} & \res{0.381}{0.184} & \res{0.00005}{0.00002} \\
0.20 & \res{0.560}{0.102} & \res{0.357}{0.176} & \res{0.00006}{0.00005} \\
0.30 & \res{0.599}{0.111} & \res{0.367}{0.212} & \bres{0.00004}{0.00002} \\
0.40 & \res{0.551}{0.150} & \res{0.306}{0.198} & \res{0.00009}{0.00010} \\
0.50 & \res{0.581}{0.136} & \bres{0.431}{0.197} & \bres{0.00004}{0.00001} \\
0.60 & \res{0.585}{0.121} & \bres{0.431}{0.186} & \res{0.00008}{0.00009} \\
\bottomrule
\end{tabular}
\end{table}

Table~\ref{tab:mask_ablation} examines sensitivity to the symmetric binary attention mask on Theia, with the other hyperparameters held fixed. The reference configuration uses a mask ratio of 0.40, selected with the same procedure as the other hyperparameters. Removing masking yields the highest mean AP, while ratios of 0.50 and 0.60 achieve the highest mean MCC at the displayed precision. The reference ratio of 0.40 has lower mean AP and MCC than the alternatives tested here. Mean validation AP ranges from 0.654 to 0.677, and test performance varies substantially across seeds, with no consistent trend as the mask ratio increases. These results do not establish a benefit from attention masking on Theia.

\subsection{Runtime Analysis}\label{app:runtime}

\begin{table}[t]
\centering
\small
\setlength{\tabcolsep}{3pt}
\caption{Theseus runtime on synthetic E3 inputs. Latency is reported as mean $\pm$ standard deviation, and memory as maximum observed process RSS and peak GPU tensor allocation.}
\label{tab:runtime_profiling}
\begin{tabular}{l
                S[table-format=4.0]
                S[table-format=5.0]
                c
                S[table-format=4.0]
                S[table-format=3.0]}
\toprule
\textbf{Dataset} &
{\makecell[c]{\textbf{Input}\\\textbf{nodes}}} &
{\makecell[c]{\textbf{Input}\\\textbf{events}}} &
{\makecell[c]{\textbf{Latency}\\\textbf{(ms)}}} &
{\makecell[c]{\textbf{CPU mem.}\\\textbf{(MiB)}}} &
{\makecell[c]{\textbf{GPU mem.}\\\textbf{(MiB)}}} \\
\midrule
Cadets   & 1200 & 8779  & \res{35.6}{10.3} & 1284 & 302 \\
FiveDir. & 4501 & 13113 & \res{94.9}{8.8} & 1339 & 746 \\
Trace    & 5504 & 13282 & \res{116.9}{9.0} & 1295 & 334 \\
Theia    & 4454 & 28339 & \res{78.1}{9.8} & 1298 & 631 \\
\bottomrule
\end{tabular}
\end{table}

We profiled Theseus on a single NVIDIA L40S using synthetic inputs based on the mean node and edge counts of the E3 snapshots in Table~\ref{tab:dataset_stats}. Graph construction adds a reverse edge for each input event and omits isolated nodes, so the resulting graph sizes differ from the input counts in Table~\ref{tab:runtime_profiling}. The table reports latency and memory usage over 30 measurements after five warmup iterations. End-to-end latency ranges from 36 to 117~ms per synthetic window, corresponding to approximately 34,000 to 57,000 input nodes per second. These measurements include synthetic embedding generation, graph construction, and inference, but not raw log parsing or model training.

To manage resource consumption, the pipeline generates 15-minute snapshots and partitions any graph exceeding 10,000 nodes into smaller subgraphs before the Transformer stage. Peak GPU tensor allocation in this runtime benchmark ranges from 302 to 746~MiB. Graph construction accounts for approximately 44\% of end-to-end latency on Theia and 62--72\% on the other datasets. Inference alone takes 9--24~ms, making graph construction a larger cost in each of these synthetic workloads.

\subsection{Hyperparameters}\label{app:hyperparameters}

Table~\ref{tab:hyperparameters} lists the final Theseus hyperparameters established under our evaluation protocol. We developed these configurations through a coarse random search with a limited budget, followed by Bayesian optimization targeting AP. The selected configurations varied with the available dataset semantics. For Cadets, FiveDirections, and Trace, explicit structural features helped offset weaker node semantics. On Theia, the model performed better without these structural additions, relying instead on populated command-line and path attributes. For ATLASv2, we used one configuration for both hosts: 300 epochs, SAGE and Transformer learning rates of $10^{-5}$ and $10^{-4}$ respectively, 0.2 dropout in both modules, one hop, a node output dimension of 96, a Transformer embedding dimension of 32, four attention heads, one Transformer layer, 15-minute windows, a batch size of 2, and structural features enabled.

\begin{table}[t]
    \centering
    \footnotesize
    \setlength{\tabcolsep}{4pt}
    \caption{Hyperparameters for Theseus.}\label{tab:hyperparameters}
    \begin{tabular}{lcccc}
        \toprule
        \textbf{Parameter} & \textbf{Cadets} & \textbf{FiveDir.} & \textbf{Theia} & \textbf{Trace} \\
        \midrule
        \multicolumn{5}{l}{\textit{Training}} \\
        Epochs & 300 & 300 & 400 & 100 \\
        Patience & 20 & 25 & 70 & 20 \\
        SAGE LR & 8.7e-5 & 6.0e-5 & 5.0e-5 & 4.8e-5 \\
        Transformer LR & 6.5e-3 & 7.0e-4 & 7.5e-4 & 7.5e-5 \\
        SAGE Weight Decay & 0.3 & 0.2 & 0.2 & 0.06 \\
        Transformer Weight Decay & 7.5e-3 & 0.05 & 0.01 & 0.07 \\
        SAGE Dropout & 0.4 & 0.3 & 0.2 & 0.25 \\
        Transformer Dropout & 0.1 & 0.2 & 0.1 & 0.2 \\
        Batch Size & 2 & 1 & 1 & 2 \\
        \midrule
        \multicolumn{5}{l}{\textit{Architecture}} \\
        Transformer Embed. Dim & 96 & 8 & 128 & 128 \\
        Node Embed. Dim & 192 & 140 & 128 & 128 \\
        Attention Heads & 2 & 4 & 8 & 8 \\
        Transformer Layers & 1 & 1 & 2 & 2 \\
        Fused Edge Counts & Yes & Yes & No & Yes \\
        Node Degrees & Yes & Yes & No & Yes \\
        \bottomrule
    \end{tabular}%
\end{table}

\clearpage
\section{Artifact Appendix}

The artifact contains the Theseus implementation, adapted evaluation code for Magic, Orthrus, and Velox, processed benchmark tables, frozen checkpoints and caches, archived result logs, and scripts for regenerating the paper's main results. Because full retraining across all systems, datasets, and seeds exceeds the one-day artifact evaluation limit, we provide a scaled workflow that uses the released checkpoints and retained preprocessing artifacts. The workflow reruns inference for every learned system, dataset, and seed included in the evaluation, regenerates the deterministic allowlist, aggregates predictions by process identity, and recomputes the corresponding metrics. Commands for full retraining remain documented in the README.

\subsection{Description \& Requirements}

\subsubsection{How to access}
Source code and documentation are available at \url{https://github.com/lorenzo9uerra/theseus}. Source commit \texttt{7477463} is archived as \texttt{theseus-source.tar.gz} at \url{https://doi.org/10.5281/zenodo.22343430}, together with caches, checkpoints, baseline artifacts, logs, and the processed ATLASv2 bundle. Processed DARPA TC E3 tables are archived at \url{https://doi.org/10.5281/zenodo.18450779}. The README identifies all files and licenses.

\subsubsection{Hardware dependencies}
Log aggregation and unit tests can run on an x86-64 Linux host with 8 CPU cores and 16~GB RAM. The semantic diagnostic requires approximately 32~GB RAM. Magic evaluation defaults to CPU (\texttt{DEVICE=-1}). Theseus and PIDSMaker inference is supported on an NVIDIA L40S, A100, or compatible CUDA GPU. The pinned PIDSMaker environment may use PTX JIT on newer architectures, adding startup time. We recommend 8 CPU cores, up to 128 GB RAM, and at least 300 GB of local storage for the complete workflow. Individual runs generally use less memory, but the largest PIDSMaker data-loading step on E3 can approach this limit. Network storage may substantially increase setup and I/O time because PIDSMaker retains many small files. Run sequentially on one GPU, the workflow takes approximately 24 hours. With two GPUs or instances, Orthrus and Velox can run concurrently, reducing expected wall time to approximately 12 hours. Full retraining may require additional storage.

\subsubsection{Software dependencies}
The tested platform is a recent x86-64 Linux distribution with an NVIDIA driver, Git, GNU Make, \texttt{tar}, \texttt{unzip}, and \texttt{uv}. Locked environments use Python 3.12--3.13 for Theseus and Python 3.9 for both baseline forks. The relevant PyTorch/CUDA dependencies are installed by \texttt{uv sync}. After downloading the archives and installing dependencies, no experiment depends on an online service or experiment tracking.

\subsubsection{Benchmarks}
The primary experiments use the processed process, file, network flow, and event tables for DARPA TC E3 Cadets, FiveDirections, Theia, and Trace, together with the released process labels and fixed temporal splits. The secondary experiment uses the packaged ATLASv2 h1/h2 Carbon Black tables and UUID-based process labels. Raw DARPA and ATLAS logs are not required. The artifact contains the exact processed inputs and labels used by the evaluation.

\subsection{Artifact Installation \& Configuration}

Download and extract \texttt{theseus-source.tar.gz} from the artifact record above. Download \texttt{DARPA-TC-E3-Parquet.zip} from the E3 record and the four \texttt{*\_artifacts.tar.gz} archives from the artifact record. From the repository root, run:
\begin{verbatim}
uv sync --frozen --extra dev
unzip -o DARPA-TC-E3-Parquet.zip -d data/
tar -xzf theseus_artifacts.tar.gz -C .
tar -xzf magic_artifacts.tar.gz \
  -C baselines/MAGIC/
tar -xzf pidsmaker_artifacts.tar.gz \
  -C baselines/PIDSMaker/
mkdir -p atlas_bundle
tar -xzf atlasv2_artifacts.tar.gz \
  -C atlas_bundle/
(cd baselines/MAGIC && make setup)
(cd baselines/PIDSMaker && make setup)
(cd baselines/PIDSMaker &&
 ln -sfn ../../data data)
(cd baselines/PIDSMaker &&
 ln -sfn ../../ground_truth ground_truth)
\end{verbatim}
\subsection{Experiment Workflow}

Each learned system loads a frozen checkpoint and retained preprocessing artifacts, reruns inference over every reported seed, aggregates repeated window outputs by unique process identity, excludes contaminated nodes, and computes process-level metrics. The deterministic allowlist is regenerated from the processed tables. Rerun logs are written to \texttt{results\_rerun/}, leaving the archived logs unchanged. Archived log aggregation should match the reference values at the precision shown in the corresponding table. Rerun inference is also expected to reproduce the rounded reference values; minor hardware-dependent numerical differences may occur, but should not alter the comparisons in C1--C4.

\subsection{Major Claims}
\begin{itemize}
  \item (C1): Under the shared process-level protocol, alerting and investigation can diverge, and on Cadets, FiveDirections, and Trace a simple executable/path allowlist matches or exceeds the learned baselines on key operating-point metrics. R0 reconstructs the complete Table~IV results, and R1 validates the comparison by rerunning inference.
  \item (C2): Theia is the only primary DARPA TC E3 dataset on which Theseus clearly improves both ranking and process recovery over the learned baselines, reaching AP 0.551, AUROC 0.994, F1 0.251, and MCC 0.306. The allowlist has slightly higher F1 and MCC. R1 reproduces these values from Table~IV.
  \item (C3): Among the primary DARPA TC E3 datasets, Theia combines the highest semantic signal quality ($Q=6.72$) with the clearest architectural separation. Cadets, FiveDirections, and Trace have $Q=1.74$, $3.97$, and $2.95$, respectively. R2 recomputes Table~V and the completeness and entropy values visualized in Fig.~3. Together, R1 and R2 support the cross-dataset comparison in C3.
  \item (C4): ATLASv2 provides mixed secondary evidence: Theseus leads AP on h1 and ADP on both hosts, whereas Velox leads AP on h2 and AUROC on both hosts. R3 reproduces Table~X.
\end{itemize}

\subsection{Evaluation}

\subsubsection{Experiment (R0)}
[Smoke Test] [5 human-minutes + 5 compute-minutes]: verifies the installation, unit tests, archived log aggregation, and deterministic allowlist.

\textit{[Preparation]}
Use the configured repository root and staged artifact archives.

\textit{[Execution]}
Run:
\begin{verbatim}
uv run pytest tests/scripts tests/tasks \
  tests/utils tests/test_main.py
uv run python scripts/aggregate_results.py
\end{verbatim}
Regenerate the deterministic allowlist using the commands under ``E3 Main Results'' in the root README, then run its aggregator. Run the baseline and ATLASv2 aggregation commands documented in the corresponding READMEs.

\textit{[Results]}
All tests pass, and each aggregator reports five runs for every learned system/dataset combination it covers. The resulting means match Tables~IV and~X after rounding.

\subsubsection{Experiment (R1)}
[Primary DARPA TC E3 Evaluation] [15 human-minutes + approximately 22 compute-hours, or approximately 12 wall-hours with two concurrent GPU jobs]: reproduces the primary result table and claims C1--C2.

\textit{[Preparation]}
Use a GPU host satisfying the hardware requirements above. One GPU supports sequential execution; when two GPU devices or instances are available, Orthrus and Velox can run concurrently.

\textit{[Execution]}
Run:
\begin{verbatim}
RESULT_DIR=results_rerun \
./scripts/reproduce_results.sh
\end{verbatim}
Run the Magic \texttt{make eval} command as documented in the root README. Run the following PIDSMaker commands sequentially on one GPU, or concurrently on separate GPU devices or instances:
\begin{verbatim}
cd baselines/PIDSMaker
make eval-orthrus RESULT_DIR=results_rerun
make eval-velox RESULT_DIR=results_rerun
cd ../..
\end{verbatim}
Aggregate each generated result directory as documented in the READMEs. On a two-GPU host, assign one visible GPU to each command and launch them in separate shells.

\textit{[Results]}
Compare the generated five-seed means with Table~IV. The results should preserve the divergence between alerting and investigation utility described in C1, and the allowlist should match or exceed the learned systems on the identified key operating-point metrics. Expected AUROC means range from 0.666 to 0.943 on FiveDirections and from 0.868 to 0.997 on Trace; Theseus should reproduce the Theia values in C2, subject to minor hardware-dependent numerical variation. ADP can vary with the ordering of tied or nearly tied scores; this sensitivity was observed for Magic on Cadets.

\subsubsection{Experiment (R2)}
[Semantic Signal Quality] [2 human-minutes + 10 compute-minutes]: reproduces the diagnostic used in claim C3.

\textit{[Preparation]}
No additional preparation is required beyond the artifact installation and configuration above.

\textit{[Execution]}
Run:
\begin{verbatim}
uv run python \
  scripts/analyze_semantic_signal_\
quality.py --all --quiet
\end{verbatim}

\textit{[Results]}
The output reports $Q=1.74$, $3.97$, $2.95$, and $6.72$ for Cadets, FiveDirections, Trace, and Theia, respectively. These results reproduce Table~V and the completeness and entropy values visualized in Fig.~3; together with R1, they support C3.

\subsubsection{Experiment (R3)}
[ATLASv2 Evaluation] [10 human-minutes + up to 1 compute-hour]: reproduces the secondary result table and claim C4.

\textit{[Preparation]}
Use the configured artifact workspace on a compatible GPU host.

\textit{[Execution]}
From the repository root, run:
\begin{verbatim}
RESULT_DIR=results_rerun/atlas \
ALLOWLIST_DIR=outputs/atlas_ae \
./scripts/reproduce_atlasv2_results.sh
\end{verbatim}

\textit{[Results]}
The generated JSON and Markdown summaries should reproduce Table~X, subject only to minor hardware-dependent numerical variation. Expected Theseus/Velox means include h1 AP 0.216/0.193, AUROC 0.911/0.918, and ADP 0.826/0.394; h2 AP 0.267/0.299 and AUROC 0.938/0.987. These results validate the mixed comparison in C4.

\end{document}